\documentclass[useAMS,usenatbib,usegraphicx]{mn2e}
\usepackage{amsmath,amsfonts,amssymb}

\newlength{\colwidth}
\newcommand{\lya}{Ly$\alpha$}

\newcommand{\1}{$^{-1}$}
\newcommand{\2}{$^{-2}$}
\newcommand{\3}{$^{-3}$}
\newcommand{\hm}{$h^{-1}$}
\newcommand{\kms}{km s\1}

\newcommand{\msun}{M$_{\sun}$}
\newcommand{\phot}{photon s\1 cm\2 sr\1}
\newcommand{\obl}{erg s\1 cm\2 sr\1 \AA\1}

\newcommand{\escma}{erg s\1 cm\2 arcsec\2}

\newcommand{\aap}{A\&A}
\newcommand{\aal}{A\&A \emph{Lett}}
\newcommand{\aj}{AJ}
\newcommand{\apj}{ApJ}
\newcommand{\apjl}{ApJL}
\newcommand{\apjs}{ApJS}
\newcommand{\mnras}{MNRAS}
\newcommand{\mnrasl}{MNRAS \emph{Lett}}
\newcommand{\pasj}{PASJ}
\newcommand{\pasp}{PASP}

\newcommand{\ssr}{SSR}

\newcommand{\spie}{SPIE}

\newcommand{\ion}[2]{#1\,{\small{#2}}}
\newcommand{\hi}{\ion{H}{I}}
\newcommand{\hii}{\ion{H}{II}}
\newcommand{\hil}{\hi~Ly$\alpha$}
\newcommand{\hilSS}{\hi~Ly$\alpha$SS}

\newcommand{\heii}{\ion{He}{II}}
\newcommand{\heiil}{\heii~Ly$\alpha$}
\newcommand{\heiih}{\heii~H$\alpha$}
\newcommand{\ciii}{\ion{C}{III}}
\newcommand{\civ}{\ion{C}{IV}}
\newcommand{\niv}{\ion{N}{IV}}
\newcommand{\nv}{\ion{N}{V}}
\newcommand{\oiv}{\ion{O}{IV}}
\newcommand{\ov}{\ion{O}{V}}
\newcommand{\ovi}{\ion{O}{VI}}

\newcommand{\neviii}{\ion{Ne}{VIII}}
\newcommand{\siiii}{\ion{Si}{III}}
\newcommand{\siiv}{\ion{Si}{IV}}

\newcommand{\cloudy}{{\sc CLOUDY}}
\newcommand{\fireball}{FIREBALL}
\newcommand{\cwi}{CWI}
\newcommand{\kcwi}{KCWI}
\newcommand{\muse}{MUSE}

\newcommand{\owls}{OWLS}
\newcommand{\default}{{\it REF}}
\newcommand{\nosn}{{\it NOSN\_NOZCOOL}}        
\newcommand{\zcool}{{\it NOZCOOL}}             
\newcommand{\wmom}{{\it WVCIRC}}               
\newcommand{\agn}{{\it AGN}}                   

\title[Rest-frame UV emission from the high-$z$ IGM]{Rest-frame UV line emission from the intergalactic medium at $2\le z\le 5$}
\author[S. Bertone and J. Schaye]
{Serena Bertone$^{1}$\thanks{E-mail: serena.bertone@gmail.com}
and Joop Schaye$^{2}$
\\
$^{1}$ Santa Cruz Institute for Particle Physics, University of California, 1156 High Street, Santa Cruz CA 95064, USA \\
$^{2}$Leiden Observatory, Leiden University, P.O. Box 9513, 2300 RA Leiden, The Netherlands }

\begin{document}
\date{Submitted to MNRAS}
\pagerange{\pageref{firstpage}--\pageref{lastpage}} \pubyear{2011}
\maketitle
\label{firstpage}

\begin{abstract}
Rest-frame ultraviolet emission lines offer the exciting possibility to directly image the gas around high-redshift galaxies with upcoming optical instruments. We use a suite of large, hydrodynamical simulations to predict the nature and detectability of emission lines from the intergalactic medium at $2\le z \le 5$. The brightest emission comes from \hil\  (1216~\AA) and the strongest metal line, \ciii\  (977), is about an order of magnitude fainter, although \hil\ may be fainter if the gas is self-shielded to the UV background or if dust is important. The highest surface brightness regions for \civ\  (1548,1551), \siiii\  (1207), \siiv\  (1393,1403) and \ovi\  (1032,1038) are fainter than the brightest \ciii\ by factors of a few. The \nv\  (1239,1243) and \neviii\  (770,780) lines, as well as \heiih\  (1640), are substantially weaker but their maximum surface brightnesses still exceed $10^2$ \phot\ at $z=2$ (for 2'' pixels). Lower ionisation lines typically arise in denser and colder gas that produces clumpier emission. The brightest \hil\ emission arises exclusively in highly overdense gas, but the highest surface brightness emission from high-ionisation metal lines traces a much wider range of overdensities. Bright metal-line emission traces gas with temperatures close to the peak of the corresponding emissivity curve. While \hil, \heiih, \ciii, \siiii, and \siiv\ are excellent probes of cold accretion flows and the colder parts of outflows, \civ, \nv, \ovi, and \neviii\ are powerful tracers of the diffuse warm-hot intergalactic medium and galactic winds. A comparison of results from simulations with varying physical prescriptions demonstrates that the predictions for the brighter metal-line emission are robust to within factors of a few. Several rest-frame UV emission lines from the high-redshift IGM will become detectable in the near future, possibly starting with the Cosmic Web Imager, which is already operating on Palomar. The Multi Unit Spectroscopic Explorer, which will be commissioned in 2012 on the Very Large Telescope, and the proposed Keck Cosmic Web Imager have the potential to revolutionise studies of the interactions between high-redshift galaxies and their environment.

\end{abstract}

\renewcommand{\ion}[2]{#1\,{\scriptsize{#2}}}


\section{Introduction}
\label{intro}

Investigating the spatial distribution of galaxies in the sky has taught us much about how the Universe evolves and structures form. However, galaxies contain only a few percent of the total baryonic mass in the Universe \citep[e.g.][]{fukugita2004} and their spatial distribution is strongly biased with respect to the dark matter distribution \citep[e.g.][]{allen2005,blanton2006}.
Mapping the distribution of the intergalactic medium (IGM) could provide one of the most powerful tools in modern astrophysical cosmology as it contains most of the baryons and is only weakly biased with respect to the dark matter. 

While absorption lines in the spectra of background quasars provide 1-dimensional information along the line of sight, emission lines map the full 3-dimensional distribution of the gas, thus providing a more realistic picture of the physical state of the gas and of its evolution with time.
Since the IGM is the main reservoir of fresh material to fuel star formation in galaxies and is in return affected by feedback \citep[e.g.][]{theuns2002,bertone2005,bertone2006,oppenheimer2008,cen2010,tescari2011}, understanding the properties of cosmic gas ultimately implies improving our understanding of the process of galaxy formation.

Predictions for the emission from the low-redshift IGM in the rest-frame ultraviolet (UV) have been made for both \hi\ \lya\  (\citealt{furlanetto2003}; \citealt{furlanetto2005}) and metal lines (\citealt{furlanetto2004}; \citealt{bertone2010b}, Paper~II hereafter). At $z>1.5$, the emission is redshifted in to the optical band and accessible to ground-based telescopes, but predictions have only been made for \hil\ and \heii\ emission \citep[e.g.][]{hogan1987,gould1996,fardal2001,furlanetto2005,cantalupo2005,yang2006,dijkstra2009,goerdt2010,kollmeier2010,faucher2010}.

\citet{cantalupo2005} and \citet{kollmeier2010} investigated fluorescent \lya\ emission induced by the UV background and by the ionising radiation of bright quasars. \citet{kollmeier2010} claim that while the detection of the glow from the diffuse IGM is beyond the sensitivity of current telescopes, quasar-induced fluorescence in high-density regions should be detectable around bright quasars \citep[see also][]{francis2004,cantalupo2005}. Indeed, such emission may already have been detected \citep{francis2006,adelberger2006,cantalupo2007}. \citet{furlanetto2005} discussed the relative importance of fluorescence and collisional excitation and claim, however, that away from bright sources the latter is the dominant channel for the production of bright \lya\ emission.
\citet{yang2006} predict that \heii\ emission at $z\approx 3$ is detectable by current facilities, and that the \heii\ 1640 \AA\ line in particular, could be a useful tool to investigate the cooling mechanisms in the so-called ``\lya\ blobs''. Recent observations by \citet{scarlata2009} and \citet{prescott2009} may have confirmed this prediction.

As of today, \lya\ blobs remain the main form of diffuse emission detected at high redshift \citep[e.g.][]{steidel2000, matsuda2004, bower2004, weidinger2005, matsuda2006, nilsson2006, saito2006, smith2007, rauch2008, yang2009, weijmans2010} but their nature is still controversial. The \lya\ emission in the blobs might be due to cooling radiation from gas infalling on to haloes \citep[e.g.][]{fardal2001, furlanetto2005, nilsson2006, dijkstra2006, smith2007, dijkstra2009, goerdt2010}, but it is difficult to rule out other mechanisms such as galactic winds driven by supernova explosions (e.g.\ \citealt{ohyama2003}; \citealt{mori2004}; \citealt{geach2009}) and particularly recombination radiation powered by internal sources \citep[e.g.][]{furlanetto2005, faucher2010}.
The detection (or non-detection) of metal-line emission from the same gas phase that produces \lya\ radiation would help reveal the nature of the blobs.

In the next few years, a number of optical integral field unit spectrographs will come online with the special purpose of detecting diffuse emission from the IGM at $z>1.5$. The Cosmic Web Imager (\cwi, \citealt{rahman2006,matuszewski2010}) started operating on the 200~inch telescope at Palomar in 2009, while the Multi Unit Spectroscopic Explorer (\muse, \citealt{bacon2009}) for the Very Large Telescope (VLT) and the Keck Cosmic Web Imager (\kcwi, \citealt{martin2009}) will begin operations in 2012/2013. These will hopefully be followed by the proposed Antarctic Cosmic Web Imager (ACWI, \citealt{moore2008}). In addition, balloon experiments in the near UV, such as the Faint Intergalactic Redshifted Emission Balloon (\fireball, \citealt{tuttle2008}), are set to explore the Universe at lower redshifts. 

These instruments have the potential to revolutionise our understanding of galaxy formation. The ability to map the surface brightness and velocity profiles of the circumgalactic gas in one or more emission lines would provide invaluable information about two crucial, but poorly understood processes: gas accretion onto galaxies and galactic winds driven by feedback from star formation and/or accreting supermassive black holes.

In this paper we investigate the rest-frame UV emission from the IGM at redshifts $2<z<5$, and explore the possibility of detecting it with ground-based, optical telescopes. We will focus on metal-line emission, but will provide some estimates for \lya\ as a point of reference.
Using a subset of simulations from the OverWhelmingly Large Simulations project (\owls, \citealt{schaye2010}), we will show that upcoming instruments such as MUSE and KCWI have the potential to detect diffuse emission from multiple transitions arising in the gas around galaxies. 

This paper is organised as follows. Section \ref{method} briefly describes the \owls\ runs and our method to calculate and map the gas emission. Section \ref{results} presents detailed results for emission in the reference (\default, hereafter) model. In Section \ref{physics} we compare the predictions of the reference model with results from a selection of other \owls\ runs with different implementations of a number of physical prescriptions.
Finally, we draw our conclusions in Section \ref{conclusion} and present convergence tests in the Appendix.

Throughout this paper we assume a WMAP3 $\Lambda$CDM cosmology \citep{spergel2007} with parameters $\Omega_{\rm m}=0.238$, $\Omega_{\rm b}=0.0418$, $\Omega_\Lambda=0.762$, $n=0.951$, and $\sigma_8=0.74$. The Hubble constant is parametrised as $H_{\rm 0} = 100$ $h^{-1}$ km s$^{-1}$ Mpc\1, with $h=0.73$. For reference, the cosmic mean density corresponds to a hydrogen number density of $n_{\rm H} = 5.1\times 10^{-6}\,{\rm cm}^{-3} [(1+z)/3]^3$. We adopt a solar abundance of $Z_{\sun} =0.0127$, corresponding to the value obtained using the default abundance set of \cloudy\  \citep{ferland1998}. This abundance set, listed in Table~\ref{table_abund}, combines the abundances of \citet{allende2001}, \citet{allende2002} and \citet{holweger2001} and assumes $n_{\rm He} / n_{\rm H} = 0.1$.

\begin{table}
\centering
\caption{Adopted solar abundances, from \citet{allende2001}, \citet{allende2002} and \citet{holweger2001}.}
\begin{tabular}{l c l c}
\hline
\hline
Element & $n_i / n_{\rm H}$ & Element & $n_i / n_{\rm H}$ \\
\hline
H  & 1                    & Mg & 3.47$\times 10^{-5}$ \\
He & 0.1                  & Si & 3.47$\times 10^{-5}$ \\
C  & 2.46$\times 10^{-4}$ & S  & 1.86$\times 10^{-5}$ \\
N  & 8.51$\times 10^{-5}$ & Ca & 2.29$\times 10^{-6}$ \\
O  & 4.90$\times 10^{-4}$ & Fe & 2.82$\times 10^{-5}$ \\
Ne & 1.00$\times 10^{-4}$ &    &                      \\
\hline
\hline
\end{tabular}
\label{table_abund}
\end{table}

\section{Numerical methods}
\label{method}

In this Section we discuss the numerical set up of this work. In particular, in Section \ref{owls} we briefly describe the \owls\ simulation project \citep{schaye2010} and in Sections \ref{gasemission} and \ref{particles} the emissivity tables and the method to calculate and map the gas emission, respectively.

\subsection{The simulations}
\label{owls}

The \owls\ project includes more than fifty different cosmological, hydrodynamical simulations with varying mass resolution, box sizes, and physical prescriptions for star formation, galactic winds driven by supernovae (SNe), feedback from active galactic nuclei (AGN), chemical evolution and gas cooling. All simulations are produced with a significantly modified version of the parallel PMTree-SPH code {\sc gadget III}, last described in \citet{springel2005}.
In this work we use a subset of simulations with a box size of 25 \hm\ comoving Mpc that contain $2\times 512^3$ particles with initial gas particle masses of $1.4\times 10^6\,h^{-1}\,$\msun\ and dark matter particle masses of $6.3\times 10^6\,h^{-1}\,$\msun. The evolution of the dark matter and gas density distribution is followed from $z=127$ down to $z=2$. 
Comoving gravitational softenings are set to $1/25$ of the mean comoving inter-particle separation, but are limited to a maximum physical scale of 0.5 \hm\ kpc, which is reached at $z=2.91$.
The box size and the mass resolution of these simulations make them ideal for studying gas on relatively large scales, but with sufficiently high resolution to resolve galaxies at $z\ge 2$.
In addition to gravitational and hydrodynamical forces, the simulations include prescriptions for radiative cooling, star formation, chemodynamics, SN and, for one simulation, AGN feedback. Below we will summarise the sub-grid physics implemented in our reference simulation. In Section~\ref{physics} we discuss a number of other models that
incorporate variations of the physical modules, implemented one at a
time. 

Radiative cooling is implemented following \citet{wiersma2009a}\footnote{Using their equation (3) rather than (4) and {\sc cloudy} version 05.07 rather than 07.02.}.
As \citet[][Paper I hereafter]{bertone2010a} and Paper~II showed, accurate cooling rates are essential to determine the intensity of the gas emission. The net radiative cooling rates are computed for 11 elements (hydrogen, helium, carbon, nitrogen, oxygen, neon, magnesium, silicon, sulphur, calcium, and iron) under the assumption of an optically thin gas in (photo-)ionisation equilibrium, in the presence of the cosmic microwave background (CMB) and the \citet{haardt2001} model for the ionising background radiation from quasars and galaxies.
The contribution of each individual element to the cooling rate is calculated by interpolating pre-computed tables (created with the publicly available photo-ionization package \cloudy, last described by \citealt{ferland1998}) as a function of density, temperature, and redshift. 

Star formation is implemented following the prescriptions of \citet{schaye2008}, according to which gas denser than $n_{\rm H,crit} \ge 0.1$ cm\3\ (i.e.\ the interstellar medium) obeys an effective equation of state of the form $P \propto \rho^{\gamma_{{\rm eff}}}$, where $P$ is the gas pressure, $\rho$ its density and $\gamma_{\rm eff}=4/3$. Gas on the effective equation of state is allowed to form stars at a pressure-dependent rate that reproduces the observed Kennicutt-Schmidt law \citep{kennicutt1998}.
Chemical evolution is implemented as described in \citet{wiersma2009b}. The simulation tracks the evolution of the abundances of the 11 elements that contribute to the radiative cooling and follows their delayed release by massive stars (Type II SNe and stellar winds) and intermediate mass stars (Type Ia SNe and asymptotic giant branch stars), assuming a Chabrier stellar initial mass function (IMF) \citep{chabrier2003} spanning the range $0.1-100$ \msun.

SN feedback is injected in kinetic form following the method of \citet{vecchia2008}. In this prescription, type II SNe inject kinetic energy locally into the gaseous neighbours of newly formed star particles. The feedback prescription includes two parameters that describe, respectively, the amount of mass injected into winds per unit stellar mass formed, $\eta$, and the initial wind velocity, $v_{\rm w}$. The \default\ model uses $\eta=2$ and $v_{\rm w}=600$ km s\1, which corresponds to about 40~percent of the total available kinetic energy from SNe, with the rest assumed to be lost radiatively.  The reference model does not include a prescription for AGN feedback, but we will consider a model with AGN in Section \ref{physics}.

\subsection{The emissivity tables}
\label{gasemission}

\begin{figure*}
\includegraphics[width=0.8\textwidth]{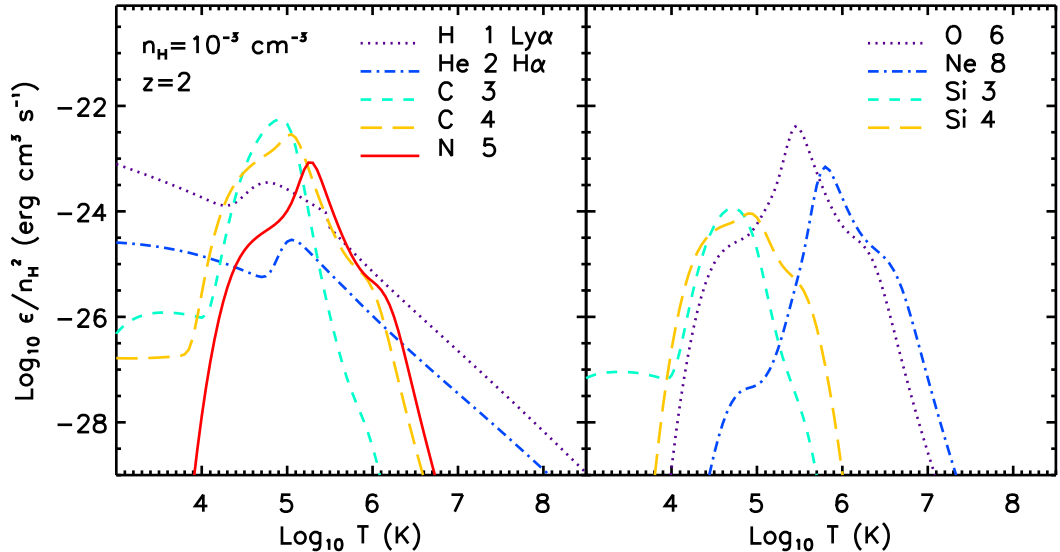}
\caption{The emissivities of a sample of emission lines as a function of temperature, for solar element abundances, hydrogen number density $n_{\rm H} = 10^{-3}$ cm\3\ and $z=2$. See Table~\protect\ref{eltable} for the corresponding rest wavelengths. The tail of the curves at the lowest temperatures is due to the effect of photo-ionisation on the ionisation balance, which is density-dependent.}
\label{z2uv}
\end{figure*}

The gas emission is calculated following the method of Paper~I and we refer the reader to that work for the details of the procedures, which we summarise briefly in the following.

Tables of the line emissivity as a function of temperature, density and redshift were created with \cloudy\  (version c07.02.02; \citealt{ferland1998}), under the assumptions of solar abundances, optically thin gas, and (photo-)ionisation equilibrium in the presence of the CMB and the \citet{haardt2001} model for the evolving UV/X-ray background radiation from galaxies and quasars.
The tables include a total of about 2000 emission lines for the 11 elements explicitly tracked by \owls.
The temperature is sampled in bins of $\Delta {\rm Log}_{10} T=0.05$ in the range $10^2\,$K $< T < 10^{8.5}\,$K and the hydrogen number density in bins of $\Delta {\rm Log}_{10} n_{\rm H}=0.2$ in the range $10^{-8}$ cm\3 $< n_{\rm H} < 10$ cm\3.

Fig.~\ref{z2uv} shows the emissivity of the lines used in this paper as a function of temperature at $z=2$, for $n_{\rm H} = 10^{-3}$ cm\3. The dependence of the line emissivity on density is illustrated in Fig.~1 of Paper II. At low temperature and in the low-density regime, the contribution of photo-ionisation by the UV background on the ionisation balance becomes important and may exceed the contribution from collisional ionisation.

The assumption of ionisation equilibrium in the calculation of the cooling rates and line emissivities is justified for regions that are predominantly photo-ionised and for dense gas in the centres of clusters, but might not be entirely robust for the warm-hot intergalactic medium (WHIM) (\citealt{yoshida2005}; \citealt{yoshikawa2006}; \citealt{cenfang2006}; \citealt{gnat2007}; \citealt{gnat2009}; \citealt{bertone2008} for a review). However, since these studies ignored photo-ionisation, they may have overestimated the importance of non-equilibrium ionisation.

The assumption that the gas is optically thin to ionising radiation may break down for very high densities. The metal lines are predominantly collisionally ionised at the temperatures corresponding to the brightest emission. This follows from the fact that the emissivity curves are very strongly peaked and that the peaks occur at temperatures $\gg 10^4\,$K, as can be seen from Fig.~\ref{z2uv}. Indeed, we will show in Section~\ref{range} that while the bright metal-line emission occurs in gas with temperatures close to the peak of the emissivity curve, the density can vary by orders of magnitudes and it is not necessarily the densest gas that dominates. For hydrogen and helium, however, the brightest emission does tend to occur in the densest gas with $T\ll 10^5\,$K. We therefore expect that self-shielding may be important for hydrogen and helium, but probably not for the metal lines.

As discussed in detail by \citet{furlanetto2005} and \citet{faucher2010}, self-shielding from ionising radiation may be very important for \hil. While a higher neutral fraction may boost the \lya\ emissivity, the absence of photo-heating may cause the temperature of the gas to drop to values that are too low to excite the electrons collisionally, thereby strongly suppressing the emissivity. As a consequence, it is possible that the \hil\ emission from the densest part of the IGM could be substantially higher or lower than we predict. Another radiative transfer effect that we will ignore is line scattering, which is important for \hil\ and can for example result in enhanced dust extinction. On the other hand, scattering of recombination radiation from \hii\ regions, which we also do not consider, may boost the \lya\ flux from the surrounding circumgalactic medium. 

Because of the potential importance of radiative transfer for \lya\ emission, we follow \citet{furlanetto2005} and consider not only the optically thin case, but also provide predictions after setting the \lya\ emissivity of all gas with $n_{\rm H} > 10^{-3}\,{\rm cm}^{-3}$ to zero. This second case is highly conservative, both because we do not expect all \lya\ photons from higher density gas to be destroyed by dust (after all, compact \lya\ emitters are known to exist) and because the gas is likely to be fully self-shielded only for higher densities, even if we consider only ionisation by the background radiation \citep[e.g.][]{Altay2011}.

We note that metal-line emission may in principle also be affected by self-shielding, although Fig.~\ref{z2uv} (see also Paper II) suggests that the bright emission is dominated by collisionally ionised gas and therefore insensitive to the amount of self-shielding. Finally, including local sources could enhance the degree of ionization and the temperature \citep[e.g.][]{schaye2006}, particularly in the case of X-ray emission \citep{cantalupo2010}. 

\subsection{The gas emission}
\label{particles}

In this Section we briefly describe the procedures to calculate the gas metal-line emission and to create emission maps. These follow the prescriptions of Paper I and we refer the interested reader to that work for more details.

The luminosities of gas particles are calculated by interpolating the emissivity tables over hydrogen number density and temperature at the desired redshift, and by rescaling the emission to the abundance of the corresponding element. The procedure is repeated for each emission line. We use the ``smoothed'' element abundances defined in \citet{wiersma2009b}, which alleviates the effect of the lack of metal mixing inherent to SPH.
The emission is calculated only for gas particles with density $n_{\rm H} \le 0.1$ cm\3, which thus excludes all particles on to which we impose an effective equation of state. We exclude these particles, which constitute the multi-phase interstellar medium in the simulations, both because we lack the resolution and physics required to model this dense gas and because we are interested in emission from the IGM. Our predictions for the surface brightness are therefore conservative as in reality the brightest emission may come from denser gas.

The luminosity in emission line $l$, in units of erg s\1, for each gas particle $i$ as a function of its gas mass $m_{{\rm gas},i}$, density $\rho_i$, hydrogen number density $n_{{\rm H},i}$ and element abundance $X_{{\rm y},i}$ is computed as
\begin{equation}\label{lumin}
L_{i,l}\left(z\right) = \varepsilon_{l,\odot} \left( z, T_i,n_{{\rm H},i} \right) \frac{m_{{\rm gas},i}}{\rho_i} \frac{X_{{\rm y},i}}{X_{\rm y \sun}},
\end{equation}
where $\varepsilon_{l,\odot} \left( z, T,n_{\rm H} \right)$ is the emissivity of the line $l$, in units of erg cm\3\ s\1\ and for solar abundances, bi-linearly interpolated (in logarithmic space) from the \cloudy\ tables as a function of the particle temperature ${\rm Log}_{10} T_{\rm i}$ and hydrogen number density ${\rm Log}_{10} n_{{\rm H},i}$ at the desired redshift $z$. $X_{{\rm y},i}$ is the mass fraction of element $y$ and $X_{\rm y \sun}$ its solar value.
The corresponding particle flux $F_{i,l}$, in units of photon s\1\ cm\2, is given by
\begin{equation}
F_{i,l} = \frac{L_{i,l}}{4\pi D_{\rm L}^2} \frac{\lambda_l}{h_{\rm P}c} \left(1+z\right),
\end{equation}
with $D_{\rm L}$ the luminosity distance, $h_{\rm P}$ the Planck constant, $c$ the speed of light and $\lambda_l$ the rest-frame wavelength of emission line $l$.

\begin{table}
\centering
\caption{Map properties. The pixel size is 2''. The angular size of the field and the velocity width of the slice both correspond to a comoving length of 25 \hm\ Mpc.}
\begin{tabular}{l l l l l}
\hline
\hline
Redshift & $2$ & $3$ & $4$ & $5$ \\
\hline
Number of pixels           & 671$^2$ & 542$^2$ & 478$^2$ & 439$^2$ \\
Pixel size (comov.\ \hm\ kpc)      & 37.2    & 46.1    & 52.3    & 56.9 \\
Angular size (arcmin)      & 22.4    & 18.1    & 15.9    & 14.6 \\
Slice thickness (\kms)     & 2234    & 2500    & 2762    & 3010 \\
\hline
\hline
\end{tabular}
\label{table_map}
\end{table}

Finally, to create emission maps we project the fluxes onto a 2-dimensional grid whose number of pixels is determined by the desired angular resolution and by the redshift we want to ``observe''. The dependence of the results on the angular resolution is illustrated in Appendix~\ref{angle}. The projection is done using a flux-conserving SPH interpolation scheme.
The gas surface brightness is computed by dividing the flux by the solid angle $\Omega$ that subtends a pixel, that is $S_{{\rm B},l} = F_{l} / \Omega$.

\begin{table*}
\centering
\caption{Summary of the emission lines considered in this work. The first column lists the ion names. The second column lists the rest wavelength $\lambda$ of the main atomic transition and the third the wavelength $\lambda_2$ of the weaker line in doublets (or Ly$\beta$ for the case of \hi). Column 4 lists the type of transition for the case of \hi\ and \heii\ and the atomic state of the ion for the others. The last four columns show the redshift range for which the transition falls within the wavelength range covered by \muse\ (4650--9300\AA) and \kcwi\ (3500--10,000\AA).}
\label{eltable}
\begin{tabular}{l r r c r r r r c}
\hline
\hline
&&&& \multicolumn{2}{c}{MUSE} & \multicolumn{2}{c}{KCWI}\\
Line & $\lambda$ (\AA) & $\lambda_2$ (\AA) & Atomic state/Transition & $z_{\rm min}$ & $z_{\rm max}$ & $z_{\rm min}$ & $z_{\rm max}$\\
\hline
\hi     & 1215.7  & 1025.7 & \lya\     & 2.8  &  6.6 &  1.9 &  7.2 \\
\heii   & 1640.0  & -      & H$\alpha$ & 1.8  &  4.7 &  1.1 &  5.1 \\
\ciii   & 977.0   & -      & Be	       & 3.8  &  8.5 &  2.6 &  9.2 \\
\civ    & 1548.2  & 1550.8 & Li	       & 2.0  &  5.0 &  1.3 &  5.5 \\
\nv     & 1238.2  & 1242.8 & Li	       & 2.8  &  6.5 &  1.8 &  7.1 \\
\ovi    & 1031.9  & 1037.6 & Li	       & 3.5  &  8.0 &  2.4 &  8.7 \\
\neviii & 770.4   & 780.3  & Li	       & 5.0  & 11.1 &  3.5 & 12.0 \\
\siiii  & 1206.5  & -      & Mg	       & 2.9  &  6.7 &  1.9 &  7.3 \\
\siiv   & 1393.8  & 1402.8 & Na	       & 2.3  &  5.7 &  1.5 &  6.2 \\
\hline
\hline
\end{tabular}
\end{table*}

Unless stated otherwise, all emission maps are created using a pixel size of 2"$\times 2$", approximately consistent with the angular resolution of \cwi\ \citep[2.5'' $\times$ 1'';][]{matuszewski2010}, and a slice thickness of 25 \hm\ comoving Mpc, equal to the size of the simulation box.
In Appendix~\ref{thick} we show that for sufficiently large fluxes, which include the regime of interest here, the probability distribution function (PDF) of the pixel flux is proportional to the slice thickness (although we do find that projection effects slightly enhance the flux for the case of \civ). For fluxes that are potentially detectable the PDF can thus easily be scaled to a different slice thickness. For simplicity, we have computed the PDFs using the entire simulation box.
At $z=2$, a slice thickness of 25 \hm\ Mpc corresponds to a spectral resolution $\Delta v = 2.2\times 10^3\,$\kms, 
which is much worse than proposed for upcoming instruments.
Even reducing the slice thickness to 1 \hm\ Mpc, as we do in Fig.~\ref{thick_pdf}, the Hubble velocity difference across the slice is larger than the resolution of \cwi.
The number of pixels in each map depends on redshift. An angular resolution of 2" requires a grid with $671^2$ pixels, which corresponds to a comoving size of 37 \hm\ kpc at $z=2$, or 74 times larger than the gravitational force resolution used in our simulations. 
For our cosmology, a region with comoving size 25 \hm\ Mpc corresponds to a field of view of about 22' at $z=2$.
The same quantities are listed for other redshifts in Table~\ref{table_map}.

\section{Results}
\label{results}

In this Section we present results for the rest-frame UV emission from the IGM in the reference model. In particular, we investigate how the intensity of the emission lines varies with redshift in Section~\ref{lines} and we  discuss the detectability in
Section~\ref{detect}. We study line ratios in Section~\ref{ratios} and we investigate how the emission depends on the properties of the gas in Section~\ref{range}. In Appendix~\ref{converge} we demonstrate that the results are converged with respect to the numerical resolution and the size of the simulation box, although it is possible that a larger box size would have allowed us to sample the PDF to higher fluxes in the regime where it is steeply declining.

\subsection{Line emission}
\label{lines}

\begin{figure*}
\includegraphics[width=\textwidth]{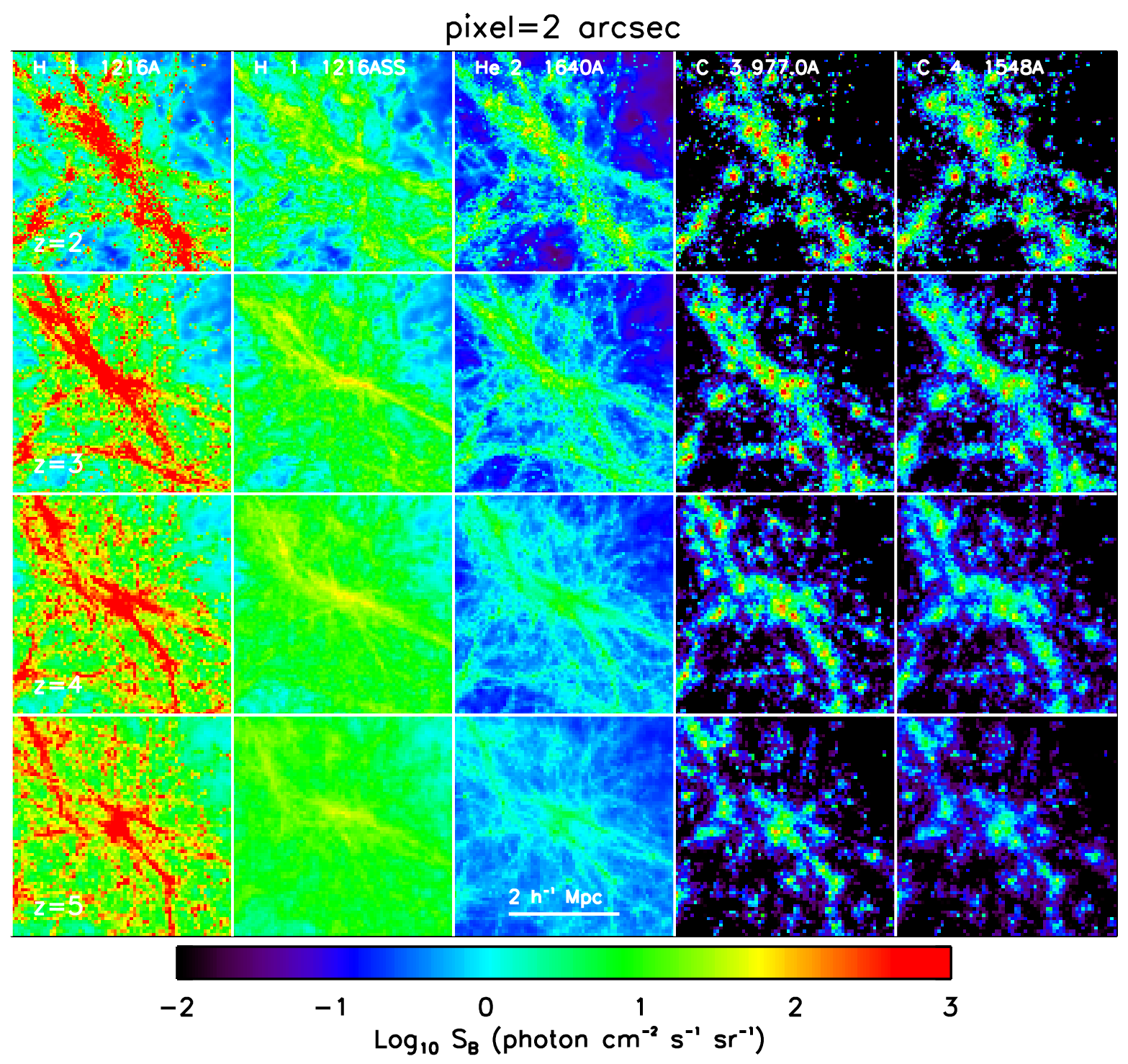}
\caption{The rest-frame UV emission of the IGM (i.e.\ gas with $n_{\rm H}<10^{-1}\,{\rm cm}^{-3}$) at redshift $z=2$ (upper panels), 3 (upper-middle panels), 4 (lower-middle panels) and 5 (bottom panels). From left to right we show emissions from \hil, \hilSS\ (i.e.\ \hil\ from gas with $n_{\rm H}<10^{-3}\,{\rm cm}^{-3}$), \heiih, \ciii, and \civ. The maps show a zoom of a region of 4 \hm\ comoving Mpc on a side that contains a group forming at the intersection of a web of filaments. All maps use a pixel size of 2", which corresponds to comoving sizes of 37, 46, 52 and 57 \hm\ kpc at $z=2, 3, 4$ and 5 respectively, and comoving slice thickness 25 \hm\ Mpc. The number of pixels shown in each panel is $108^2$, $87^2$, $76^2$ and $69^2$ at $z=2, 3, 4$ and 5, respectively. A given colour corresponds to the same surface brightness level in all panels. The minimum and maximum values used for the colour scale do not correspond to the actual minimum and maximum.}
\label{allmaps}
\end{figure*}

\begin{figure*}
\includegraphics[width=\textwidth]{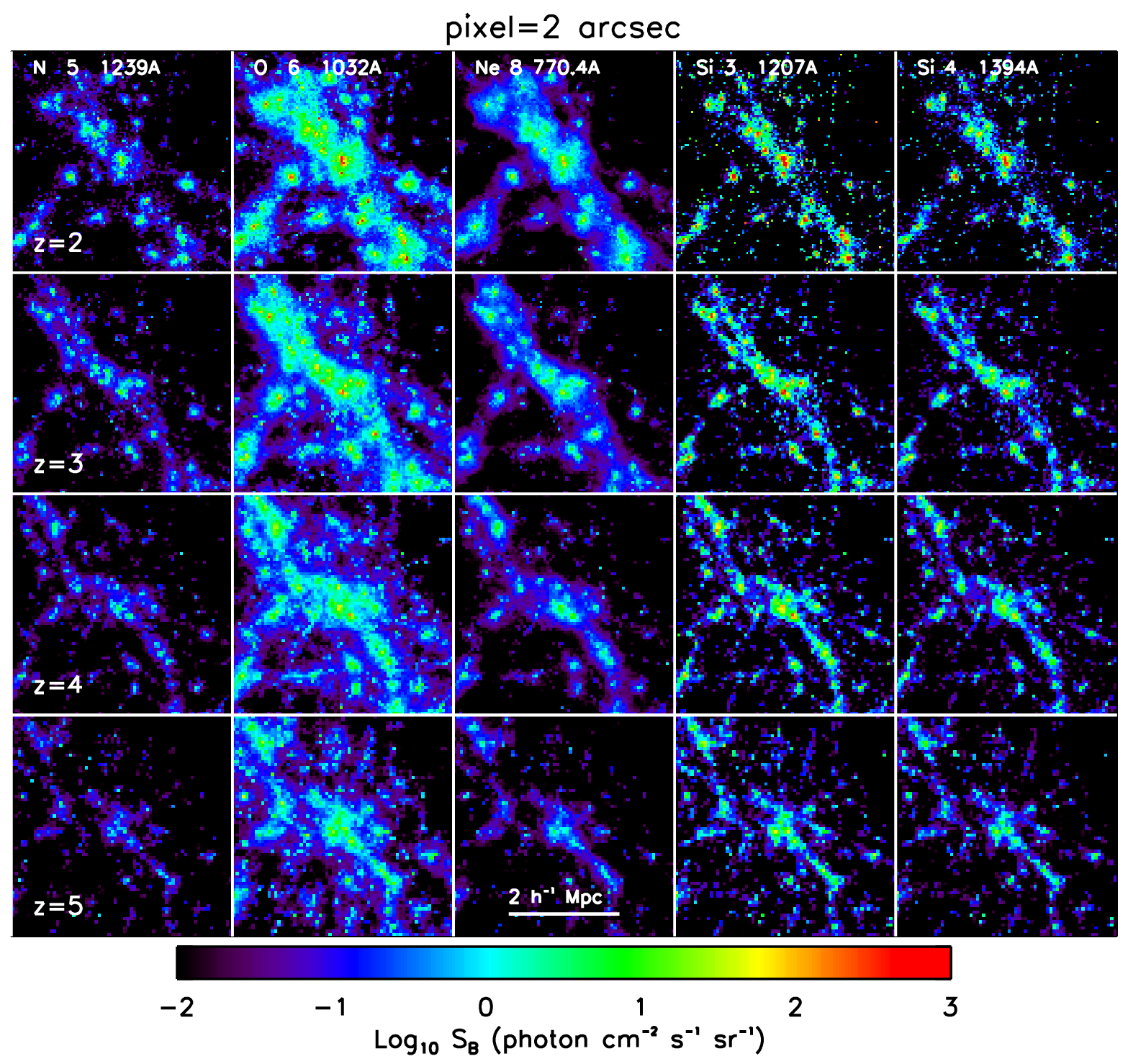}
\caption{Same as Fig.~\ref{allmaps}, but for \nv, \ovi, \neviii, \siiii, and \siiv, from left to right.}
\label{othermaps}
\end{figure*}

\begin{figure*}
\includegraphics[height=0.95\textwidth,angle=90]{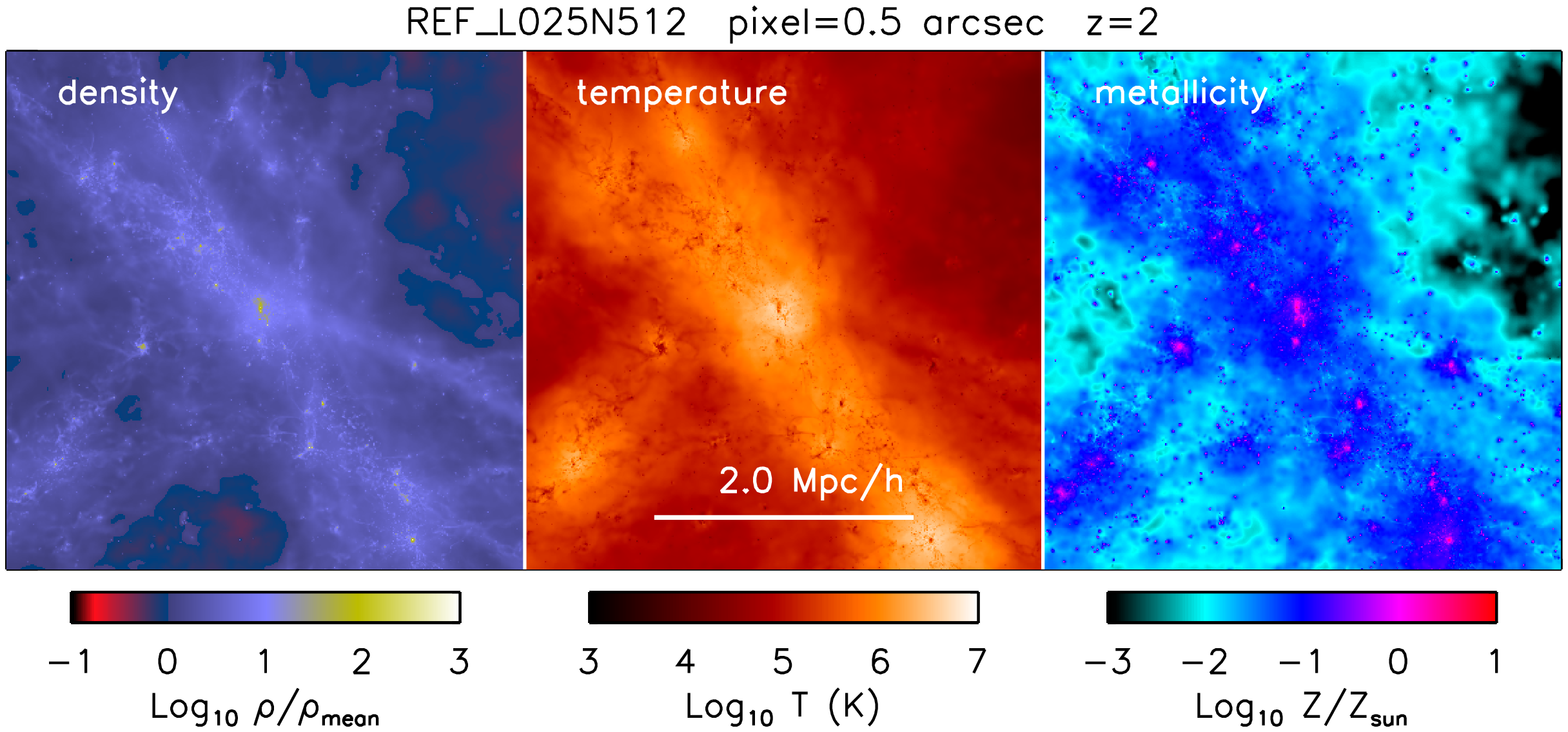}
\caption{Maps of the density (left panel), temperature (middle panel) and metallicity (right panel) of the gas in the \default\ simulation at $z=2$. The displayed region is the same as that shown in Figs.~\ref{allmaps} and \ref{othermaps}. It shows a zoom of a region of 4 \hm\ comoving Mpc on the side that contains a group forming at the intersection of a web of filaments. All maps use a pixel size of 0.5", which corresponds to a comoving size of 9 \hm\ kpc at $z=2$, and slice thickness 25 \hm\ comoving Mpc.}
\label{maps2}
\end{figure*}

In this Section we investigate the strengths of the emission lines listed in Table~\ref{eltable} and their evolution from $z=5$ to 2. The lines have been selected on the basis of their strength or observability in the optical band. The minimum and maximum redshifts for which the various lines fall within the wavelength ranges of \muse\ and \kcwi\ are listed in Table~\ref{eltable}.  

Among the promising metal lines are those from lithium-like ions, i.e. \civ\  (1548,1551), \nv\  (1238,1243), \ovi\  (1032,1038) and \neviii\  (770,780), as well as \ciii\  (977), \siiii\  (1207), and the sodium-like ion \siiv\  (1394,1403). We will also present results for two important cooling lines of hydrogen and helium: \hil\  (1216) and \heiih\  (1640). Several other rest-frame UV lines, e.g.\ \heiil\  (304), \niv\  (765), \oiv\  (549), and \ov\  (630), are usually about an order of magnitude stronger than lines from more ionised atoms. However, we do not consider them here because they only shift into the optical at very high redshifts.

The lines that are singlets may be hard to identify unless they are cross-correlated with other lines and/or galaxies. However, these lines are interesting for a number of reasons. Firstly, they provide a sensitive diagnostic for gas at lower temperatures than the standard lines from lithium-like atoms. Secondly, the simultaneous detection of lines from different ions provides stronger constraints on the ionisation state of the emitting gas, and therefore on its density and temperature. Finally, we will show in the following that, in the
absence of strong foreground absorption, the singlets \ciii\ and \siiii\ are stronger than the corresponding doublets \civ\ and \siiv.

Figs.~\ref{allmaps} and \ref{othermaps} show maps of the surface brightness of all lines in a high-density region centered on a forming group. The line intensity is mapped at $z=2, 3, 4$ and 5, for a pixel size of 2''. In case of doublets, such as for example \civ\ and \ovi, we have only
considered the strongest component. The intensity of the weaker
line in the doublet can be obtained by dividing the intensity of the
stronger one by a factor of 2. Note that not all lines fall within
the optical range at all redshifts shown. For comparison, maps of the density, temperature and metallicity of the gas in the same region are
shown in Fig.~\ref{maps2} at $z=2$ for a pixel size of 0.5". For reference, 
\begin{eqnarray}
&& 10^{-19}\,{\rm erg}\,\,{\rm s}^{-1}\,{\rm cm}^{-2}\,{\rm arcsec}^{-2} \approx \nonumber \\
&& 8.57 \times 10^2  \left( \frac{\lambda_0}{1000\, \textrm{\AA}}\right)
\left( \frac{1+z}{4} \right)
{\rm photon~s}^{-1} {\rm cm}^{-2} {\rm sr}^{-1}
\label{eq:unitconv}
\end{eqnarray}
where $\lambda_0$ the rest-frame wavelength.

We showed in Paper~II that at low redshift the UV emission comes from dense gas in filaments and in the haloes of galaxies, but not from the centres of large groups, because these contain gas heated to temperatures in excess of $10^6\,$K and emit mostly in the X-ray band. At high redshift, the situation is different. At $z>2$ massive structures such as clusters and large groups have not yet fully assembled. Moreover, fixed overdensities correspond to higher proper densities and thus higher cooling rates. Cold streams are therefore more prominent at higher redshift.
Most of the UV emission comes from gas inside or in the vicinity of collapsing objects, such as the haloes of galaxies and the gas that starts to accumulate in filaments and groups. Little or no metal-line emission is produced by the lowest density IGM.

\begin{figure*}
\includegraphics[width=0.8\textwidth]{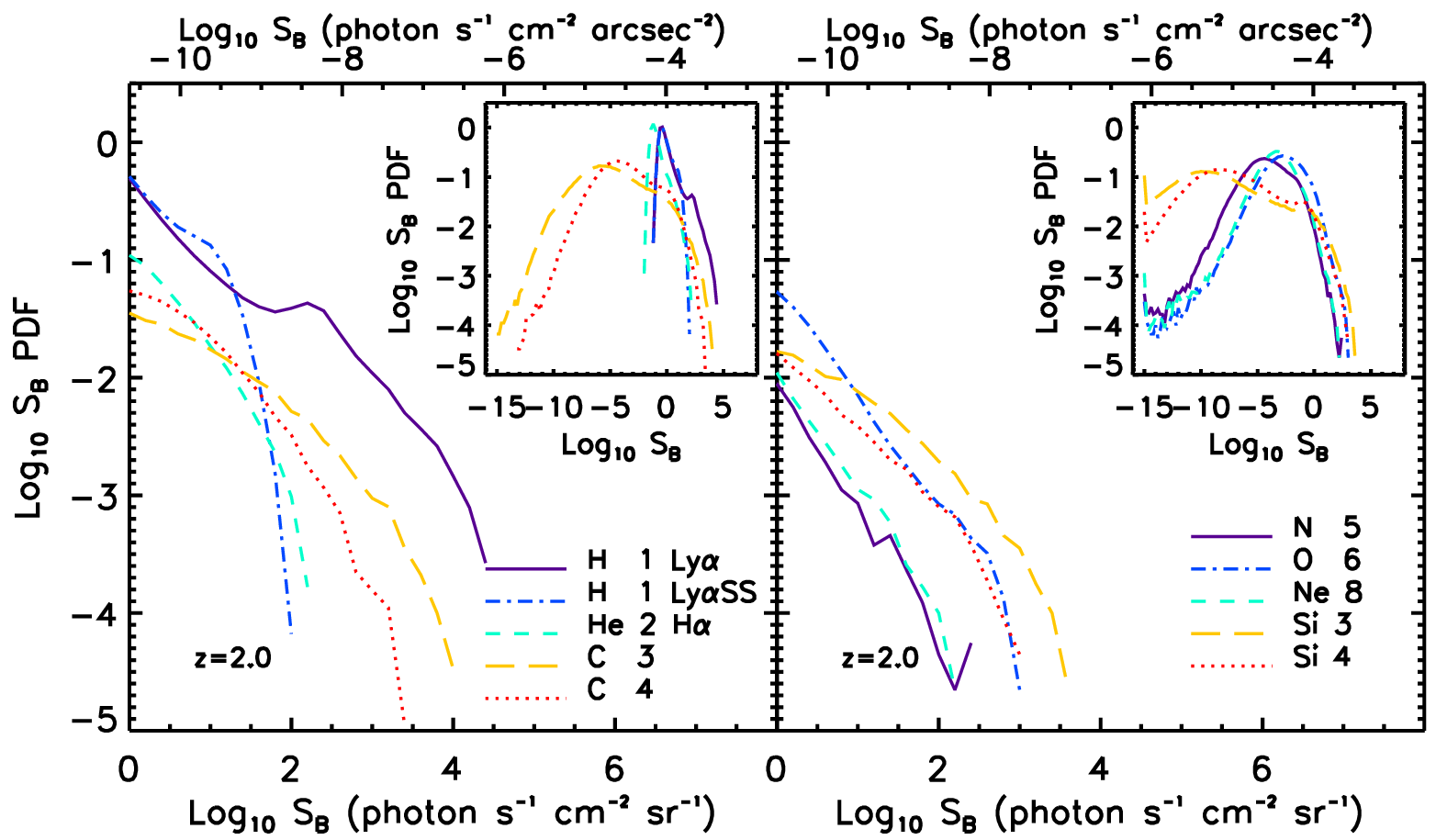}
\caption{The $z=2$ surface brightness PDFs for emission from the $z=2$ IGM (i.e.\ gas with $n_{\rm H}<10^{-1}\,{\rm cm}^{-3}$) for our sample of rest-frame UV emission lines, also shown in Figs.~\ref{allmaps} and \ref{othermaps} and listed in Table~\ref{eltable}.
The left panel shows results for \hil, \hilSS, \heiih, \ciii, and \civ, and the right panel for \nv, \ovi, \neviii, \siiii, and \siiv.
The main plotting area shows only the high flux tail of the distributions, while the full PDFs are shown in the inset. The pixel size is 2" (which corresponds to a comoving size of 37~\hm~kpc) and the PDFs are computed from maps covering the entire simulation box. In Appendix~\ref{thick} we show that for sufficiently large fluxes ($>0.1$~\phot\ for \ovi) the PDF is proportional to the slice thickness, which we have taken to be 25~\hm\ comoving Mpc (i.e.\ the size of the simulation box).}
\label{lines_pdf}
\end{figure*}

Fig.~\ref{lines_pdf} shows the surface brightness probability distribution function (PDF) of emission lines produced by the IGM (i.e.\ $n_{\rm H}<10^{-1}\,{\rm cm}^{-3}$) at $z=2$. The left panel shows results for \hil, \hilSS\ (i.e.\ \hil\ from gas with $n_{\rm H} < 10^{-3}\,{\rm cm}^{-3}$, as would be appropriate if no \lya\ photons were produced in or escaped from self-shielded gas), \heiih, \ciii, and \civ\ and the right panel for \nv, \ovi, \neviii, \siiii\ and \siiv. In all panels the plot in the main window shows the high flux tail of the PDF, while the inset shows the full distribution. For numerical convenience, the minimum value of the surface brightness has been fixed to $S_{\rm B}=10^{-15}$ \phot. This is reflected in the high values of the PDFs of metal lines for the lowest values in the inset plots. Note that these PDFs, as well as the ones in other figures, assume 2'' pixels and that, different from Figs~\ref{allmaps} -- \ref{maps2}, they have been computed from maps covering the entire simulation box. The dependence on the pixel size is illustrated in Fig.~\ref{angle_figure} for a subset of the lines.   

\begin{figure*}
\includegraphics[width=\textwidth]{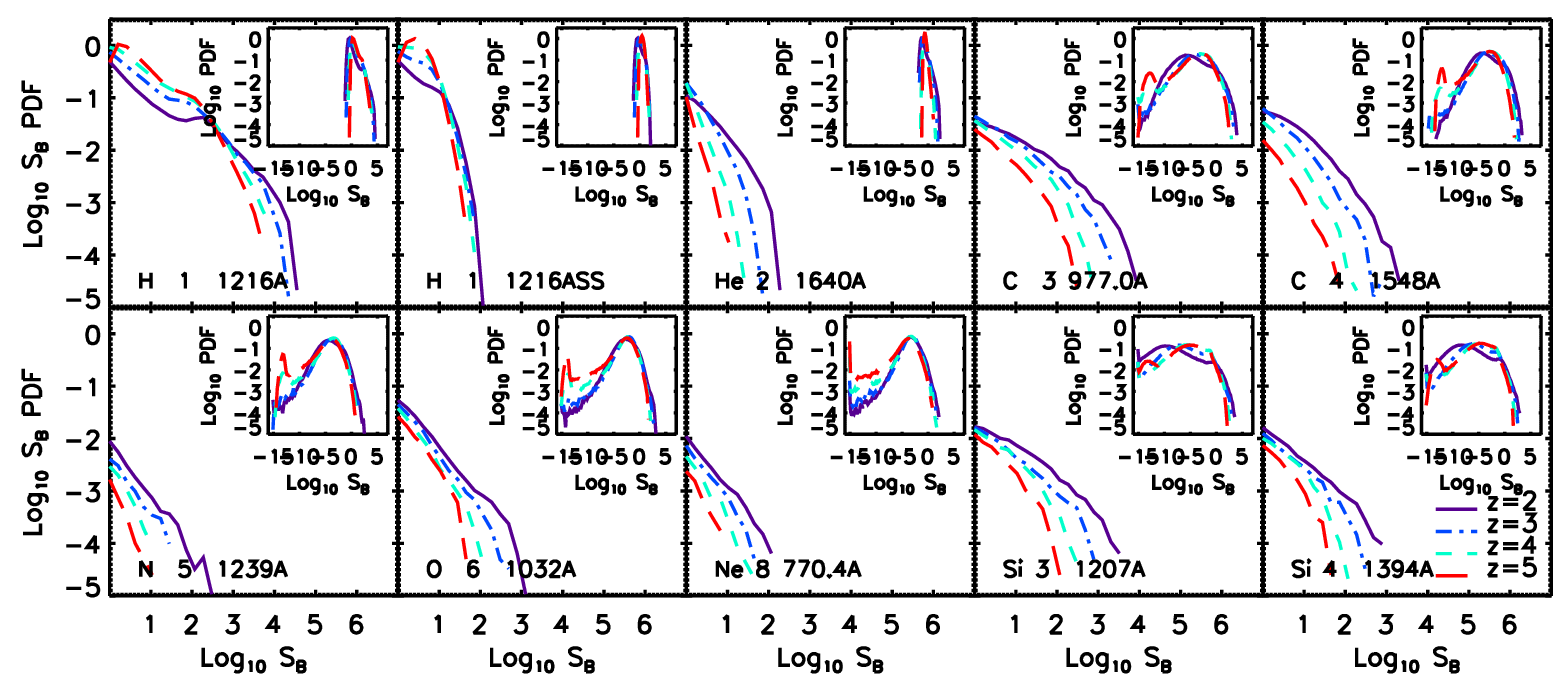}
\caption{As Fig.~\ref{lines_pdf}, but for \hil, \hilSS, \heiih, \ciii, and \civ\ (upper panels, from left to right), \nv, \ovi, \neviii, \siiii, and \siiv\ (bottom panels). Each panel shows the results for $z=2, 3, 4$ and 5. The surface brightness on the $x$-axis is in units of \phot.}
\label{redshift_pdf}
\end{figure*}

Fig.~\ref{redshift_pdf} shows the evolution of the flux PDF for individual emission lines, comparing the results at $z=2,3,4$ and 5.
Figs.~\ref{lines_pdf} and \ref{redshift_pdf} demonstrate that if we assume the gas to be optically thin, then the strongest rest-frame UV emission line among those shown here is \hil, which at $z=2$ has an intensity that is about an order of magnitude greater than the brightest \ciii, which is itself a factor of a few brighter than the strongest \civ. The difference becomes even larger at higher redshifts, where the emission from metals is suppressed because of the lower gas metallicity. If, instead, we conservatively assume that no \hil\ comes from gas with $n_{\rm H}>10^{-3}\,{\rm cm}^{-3}$ (case \hilSS), then the bright-end of the flux PDF is cut off at $\sim 10^2\,$\phot\ and the strongest emission is dominated by metal lines.
From the metal lines shown here, the brightest emission is produced by \civ, \siiv\ and \ovi\ among the higher ionisation states and by \ciii\ and \siiii\ among the lower ionisation states.

From Fig.~\ref{allmaps} we see that \ciii\ and \civ\ trace each other relatively well, with \ciii\ somewhat stronger and more biased to higher density gas. The same is true for \siiii\ and \siiv\  (Fig.~\ref{othermaps}), which, however, fall off more rapidly with declining density than the carbon lines. The other metal lines, and \ovi\ in particular, are distributed more smoothly. The maps illustrate clearly that lines from higher ionisation states are better tracers of lower density regions like filaments. In practice, the emission of all lines outside collapsed regions is extremely low and well below the direct detection limits of planned detectors.

The intensities of \nv\ and \neviii\ lines are lower than those of oxygen and carbon by at least an order of magnitude. \nv\ might be barely strong enough to be detectable at $z\approx 2$. While \neviii\ lines are unlikely to be detectable by \cwi, they might be within reach of \kcwi. However, at $z<3.5$ a near UV instrument is required.

Fig.~\ref{redshift_pdf} shows that the maximum value of the predicted flux increases with decreasing redshift for all lines. For example, at $z=5$ the maximum \ovi\ intensity is about 10 \phot, while it is about 300 \phot\ at $z=2$, i.e.\ more than an order of magnitude larger. At the highest redshift a significant number of pixels in the map do not contain any metal-line emission, as the spikes in the PDFs at the minimum value $10^{-15}$ \phot\ demonstrate. However, as time proceeds, the fraction of pixels with very low or no metal-line emission strongly decreases, indicating the progression of the IGM metal enrichment.
The weak evolution of the intensity of the \lya\ radiation, with variations well within an order of magnitude, is in good agreement with \citet{furlanetto2005}. 

The shape of the flux PDF varies for different lines and changes with time.
The PDFs of the \ovi\ and \neviii\ lines are narrow and peak at $>10^{-5}$ \phot\ at all redshifts. For most other lines, and particularly for \ciii, \siiii, and \siiv, the peak of the distribution moves to low fluxes ($10^{-10} - 10^{-5}$ \phot) and the distribution itself tends to flatten out as redshift decreases.
The behaviour of the PDFs of the \civ\ and \nv\ lines is somewhat intermediate between these two cases.
This can be explained by the fact that the emission lines trace different gas phases with different spatial distributions, which ultimately are responsible for the shape of the PDF.
As we will discuss in more detail in the next Sections, \ovi\ and \neviii\ lines trace warm-hot gas with $T\sim 10^6\,$K, while most other lines trace warm gas with $T\la 10^5\,$K. The fraction of the volume filled by shock-heated gas at $T\sim 10^6\,$K increases with time, while the metal-enriched, warm gas with $T<10^5\,$K that produces \siiii, \siiv, \ciii\ and \civ\ emission becomes progressively more confined to the haloes of galaxies and small groups as the Universe expands. As a consequence, the PDFs of these lines flatten and move toward lower fluxes.
The behaviour of \nv\ lines, produced by gas at $T\sim 10^5\,$K, is intermediate between these two cases.

\subsection{Detectability}
\label{detect}

\begin{table*}
\centering
\caption{The characteristics of the \fireball\ balloon experiment
\citep{tuttle2008} as well as the \cwi\ \citep{matuszewski2010}, \kcwi\ \citep{martin2009}, and \muse\ \citep{bacon2009} instruments. }
\begin{tabular}{l r c c c c c}
\hline
\hline
Instrument & Mirror diam. & Wavelength range & Bandwidth & FoV & Angular resolution & Resolving power\\
           & (m) & (\AA ) & (\AA) & (arcsec$^2$) & (arcsec) \\
\hline
\fireball  & 1  & $1970-2130$  & 160 & $160\times 160$ & 10.0  & 5000 \\
\cwi       & 5  & $3800-9500$  & 150 & $60\times 40$  & $2.5\times 1$  & $\ge 5000$ \\
\kcwi      & 10 & $3500-10000$ & $300 - 2500$ & $20\times (8.4-34)$ & $0.35-1.4$ & $900-20000$ \\
\muse      & 8  & $4650-9300$  & 4650 & $60\times 60$  & 0.2  & $1750-3750$ \\
\hline
\hline
\end{tabular}
\label{exper}
\end{table*}

A number of optical instruments will come online within the next few years with the special purpose of detecting diffuse emission from the IGM at $z>1.5$. In 2009 \cwi\ \citep{rahman2006,matuszewski2010} started operating on the 200~inch telescope at Palomar. \muse\ \citep{bacon2009} will be commissioned at the VLT in 2012 and \kcwi\ \citep{martin2009} will hopefully be installed at Keck soon after that. In the more distant future these instruments may be followed by the proposed ACWI \citep{moore2008}, which would operate from Antarctica. In addition, the exploration of the low-redshift Universe has begun with the near UV balloon experiment \fireball\ \citep{tuttle2008}.

The main characteristics of \cwi, \muse, \kcwi\ and \fireball\ are listed in Table~\ref{exper}. \fireball\ only covers a narrow wavelength range in the UV, $1970-2130$\AA, but has a large FoV of 160'' $\times$ 160'' (angular resolution 10''). \cwi\ has a 60''$\times 40$'' FoV sampled by 24 2.5''$\times 40$'' reflective slits. Although the wavelength range is large\footnote{Currently CWI's band width is still limited to $4400-5600\AA$ \citep{matuszewski2010}.}, the instantaneous bandwidth is limited to 150\AA. The
\kcwi\ will have various modes. The highest spectral resolution mode, $R\approx 20,000$, will only be available with a small FoV of 20''$\times 8.4$'' (angular resolution 0.35'') and for a narrow, instantaneous band pass of 300\AA. On the other hand, at the coarsest spectral resolution, $R\approx 900$, the FoV is 20''$\times 34$'' (angular resolution of 1.4'') and the band width 2500\AA. There will be a number of intermediate settings as well. While \muse\ has a much bigger FoV (60''$\times$ 60''), a higher angular resolution (0.2'', note that \muse\ will make use of adaptive optics) and a larger instantaneous band width ($4650-9300$\AA), its maximum resolving power is limited to $R=3750$ and it will not cover the blue. 

For a spectrally unresolved emission line,\footnote{The spectral resolution of \muse\ varies with wavelength but is typically $\sim 10^2\,$\kms. Using a slice thickness that matches the spectral resolution does not change our conclusions for metal lines. For \hil\ the situation may, however, be different because resonant scattering, which our simulations do not model, may strongly broaden the lines.} an 80 hour observation with \muse\ will reach a limiting surface brightness of $2\times 10^{-19}\,$\escma\ for a S/N of five and a 1 arcsec feature at $\lambda=5000$~\AA. We expect \kcwi\ to reach similar depths. Using equation (\ref{eq:unitconv}), we see that \muse\ and \kcwi\ should be able to detect intensities of $\sim 2 \times 10^3\,$\phot\ from the ground for 1'' features and a S/N of five. For more extended emission fainter levels could be reached, of course all provided that the background can be subtracted to the required accuracy. As some of the emission will indeed be
extended (see Fig.~\ref{angle_figure}), we thus expect deep
observations with upcoming instruments to be able to directly detect
diffuse emission down to $\sim 10^2 - 10^3$ \phot. Stacking galaxies may allow fainter emission to be detected statistically. Indeed, 
\citet[][]{steidel2011} have already detected diffuse \lya\ emission down to $10^{-19}\,$\escma\ by stacking $z\sim 2.65$ star-forming galaxies.

Table~\ref{eltable} lists the minimum and maximum redshifts for which
the various lines that we investigate fall within the range of
\muse\ and \kcwi, respectively. The redshift range for \cwi\ is only
slightly smaller than that of \kcwi\ (see Table~\ref{exper}).

According to our predictions, a number of emission lines should be detectable by upcoming instruments. As a benchmark, we use the instrument sensitivity of 200 \phot, reported by \citet{rahman2006} for CWI for 10~arcsec features. We note, however, that \citet{matuszewski2010} quote a much worse limiting sensitivity of $10^{-18}\,$\escma\ for features of 600~arcsec$^2$ and a 16 hour integration, which corresponds to $\sim 10^4\,$\phot. They note, though, that the sensitivity can be improved by over an order of magnitude if the background subtraction is improved using the nod-and-shuffle technique. In any case, 200 \phot\ will for example be reachable with \muse\ for $\sim 10$'', spectroscopically unresolved features (at 5$\sigma$ in an 80 hour integration).

\hil\ emission from regions with $\rho > 10^3 \bar{\rho}_{\rm b}$ should be detectable up to at least $z=5$ (taking our optically thin calculations at face value), with the glow of less dense regions ($\rho \sim 10^3 \bar{\rho}_{\rm b}$) visible only at lower redshifts. The \heiih\ line may be visible in dense (but still $n_{\rm H} < 10^{-1}\,{\rm cm}^{-3}$) gas at the lowest redshift we consider ($z\sim 2$). \civ, \ovi\ and \siiv\ will all be detectable up to $z=3$, while \ciii\ and \siiii\ should be detectable up to $z\approx 4$ or higher. The \nv\ and \neviii\ lines should also be within the capabilities of \muse\ and \kcwi\ and with some luck they might even be able to detect weak \oiv\ and \ov\ emission at $z>4.5$. In practice, however, detection of \ovi, \ov\ and \neviii, which have rest-frame wavelengths shorter than the hydrogen Lyman limit (912~\AA), may not be feasible due to the high optical depth provided by intervening absorption systems. The same may be true for \ciii\ and \siiii, which have rest-frame wavelengths of 977.0 and 1206.5~\AA and are thus susceptible to intervening \lya\ absorption. As we noted earlier, statistical detections, e.g.\ by stacking \hil\ emitters, should enable the detection of fainter lines. 

Recall that since our focus is the IGM, we excluded all gas with densities $n_{\rm H} > 0.1~{\rm cm}^{-3}$. As the emissivity scales as density squared, this high-density gas may be more easily detected, particularly for instruments with high angular resolution such as \muse\ and \kcwi. On the other hand, for observations with instruments with relatively poor angular resolution, such as \cwi, it may be difficult to tell whether apparently diffuse emission is in fact due to a few unresolved, dense knots, thus potentially mistaking emission from the ISM of small galaxies for emission from diffuse, intergalactic gas. 

\begin{figure*}
\includegraphics[width=\textwidth]{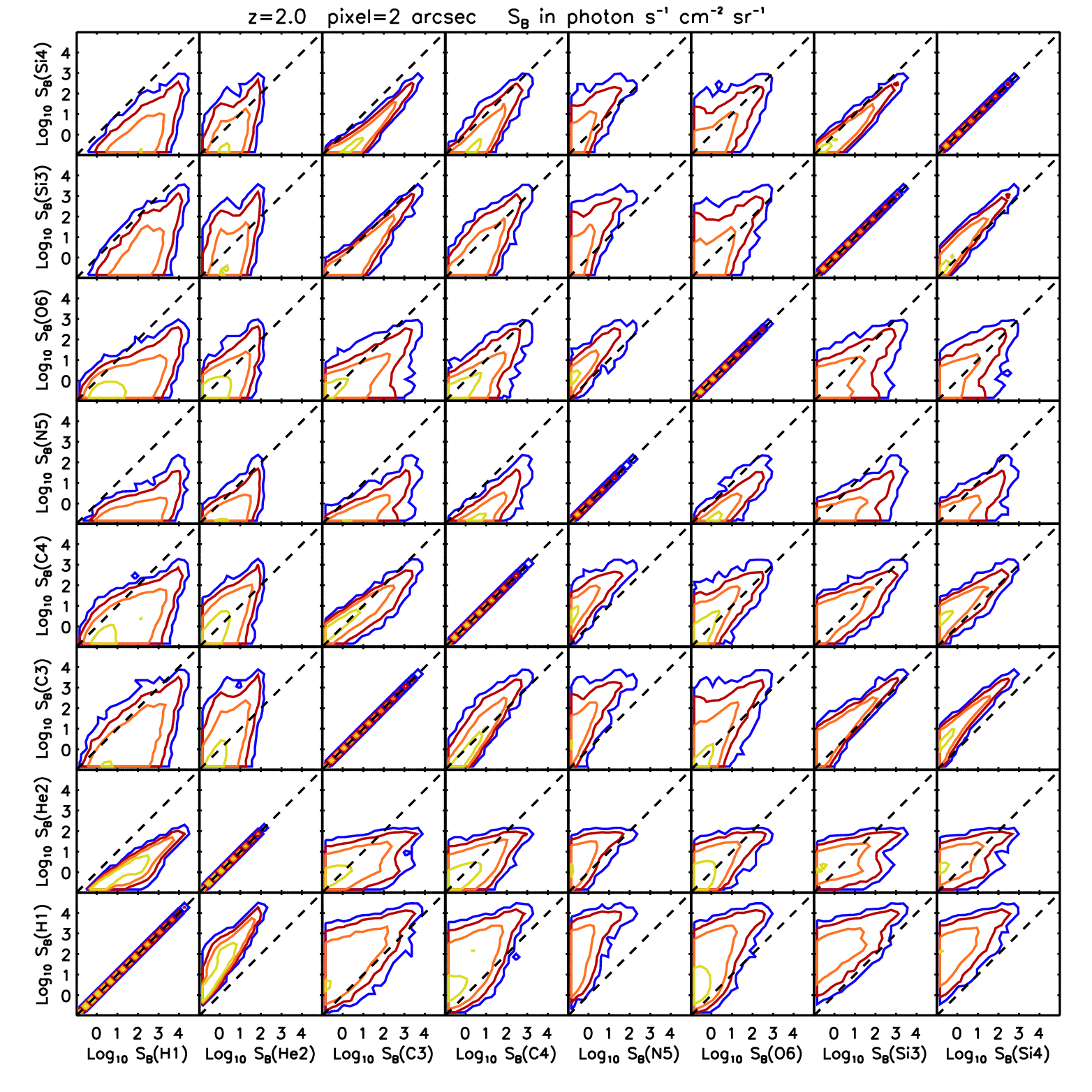}
\caption{Comparison of line intensities for a subsample of emission lines, namely \hil, \heiih, \ciii, \civ, \nv, \ovi, \siiii, and \siiv. The contours are spaced logarithmically by 1 dex. The distribution of line ratios is shown only for the high surface brightness tail of the PDF, i.e.\ intensities higher than 0.1 \phot. The correlation is roughly linear for lines from the same atom (\ciii\ and \civ, \siiii\ and \siiv). The scatter in the distributions varies from ratio to ratio, and is largest for lines with emissivities which peak at very different temperatures.}
\label{flux}
\end{figure*}

It will be necessary to reach fluxes well below the optical background light (OBL, hereafter) to detect emission from the diffuse IGM. Most of the OBL luminosity is due to zodiacal light, with only relatively minor contributions from the extragalactic background light and the diffuse galactic emission. The intensity of the OBL has been measured to be $33.5\times 10^{-9}$, $105.7\times
10^{-9}$, and $72.4\times 10^{-9}$ \obl\ at 3000, 5500, and 8000 \AA, respectively (\citealt{bernstein2007}; see also \citealt{bernstein2002} and references therein). Assuming a resolving power of $R=\lambda/\Delta\lambda =10,000$, these correspond to $\sim 2\times 10^3$, $2\times 10^4$ and $2\times 10^4$ \phot\ at 3500, 5500 and 8000 \AA, respectively.

As discussed in Paper II, the \fireball\ balloon experiment \citep{tuttle2008} provides a good chance to detect IGM emission in the UV band at lower redshifts and shorter wavelengths than \cwi. The instrument targets the wavelength window 1970-2130 \AA, where atmospheric absorption is minimal. The surface brightness limit is about 2000 \phot\ for an angular resolution of 10". The instrument aims to detect \hil, \civ\ and \ovi\ emission at $z\lesssim 1$. According to our predictions, in this energy range it might be possible to detect the \ov\ line at 630 \AA\ at $z\approx 2.2$, if intervening absorption is not very strong. For this emission line at this redshift we predict a maximum intensity $>10^3$ \phot, which might be within reach of the instrument capabilities. \fireball\ might also be able to detect \ciii\ emission at $z\approx 1$, whose surface brightness should be about an order of magnitude larger than its detection limit. Unfortunately, both the \ov\ and the \ciii\ lines are singlets and might thus be difficult to identify in spectra with limited redshift and wavelength coverage. As such, it is likely that both lines will simply be treated as unknown contaminants. However, cross-correlation with stronger lines, such as \hil, may allow a statistical detection of singlets.
A positive detection would provide proof that \ciii\ and \ov, and lines from beryllium-like atoms in general, are indeed an efficient cooling channel and would encourage further studies of emission in the far UV band.

Finally, we caution the reader that the discussion presented in this section should only be taken as a rough guide to what may be possible with upcoming facilities. More robust conclusions would require a proper study of detectability, which would involve analysing virtual observations targeted to specific instruments, instrumental modes, and observing strategies and using realistic analysis techniques. While such a study is clearly beyond the scope of the present work, the results presented here suggest that such investigations are worthwhile for several transitions and redshifts.

\subsection{Line ratios}
\label{ratios}

In this Section we compare the intensity of pairs of lines. Results for a subsample of emission lines at $z=2$ are presented in Fig.~\ref{flux} for the \default\ model. The chosen lines are representative of all interesting transitions. Fig.~\ref{flux} focuses only on the high surface brightness end of the distribution, with $S_{\rm B} \ge 10^{-1}$ \phot.

As we will discuss further in the next Sections, the intensity of the emission lines depends on the gas temperature, density and metallicity. At a fixed temperature, sufficiently high for collisional ionisation to dominate over photo-ionisation, the relative intensity of one line with respect to another is independent of the density and proportional to the ratio of the elemental abundances. Since the shape and values of the emissivity curves as a function of temperature are characteristic for each line, the line ratios vary line by line.
This is seen in Fig.~\ref{flux}. The lines of different ions of the same element are strongly correlated, but the line ratios of ions from different elements show a larger scatter.  This is because the gas that produces strong emission in one line is often different from gas that produces emission in a different line. Indeed, the scatter is largest for ratios between lines whose emissivities peak at very different temperatures. \hil\ emission is almost always stronger than any other line (but note that this changes if we ignore \hil\ from self-shielded gas; see e.g.\ Fig.~\ref{lines_pdf}) , with the exception of a few pixels in which carbon and oxygen lines dominate. These likely trace highly metal enriched gas in the haloes of galaxies and groups.

The ratio between the intensity of pairs of emission lines ultimately provides information about the ionisation state and the temperature of the gas. The intensity of a line is highest when the gas has a temperature close to the peak temperature of its emissivity curve. A higher value of the line ratio indicates that the gas temperature is closer to the peak temperature of the line in the numerator.

\subsection{What gas produces the emission?}
\label{range}

\begin{figure}
\includegraphics[width=\colwidth]{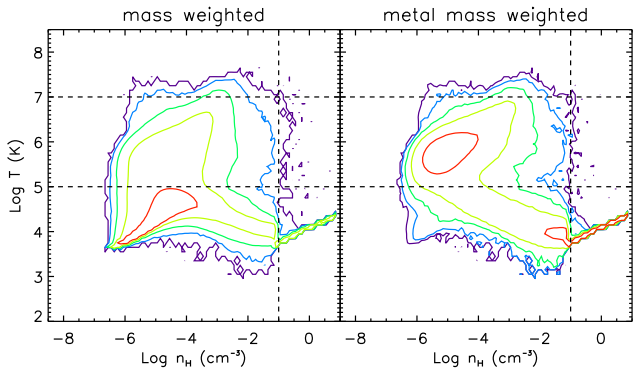}
\caption{The distributions of gas (left panel) and metal mass (right panel) in the temperature-density plane at $z=2$. Results are for the \default\ simulation. The contours are spaced logarithmically by 1.5 dex. The vertical line indicates the threshold at $n_{\rm H} = 0.1$ cm\3\ above which we impose an effective equation of state for star-forming gas \citep{schaye2008}. The horizontal lines define the range of temperatures of WHIM gas, $10^5\,$K $<T<10^7\,$K.}
\label{twomass}
\end{figure}

\begin{figure*}
\includegraphics[width=\textwidth]{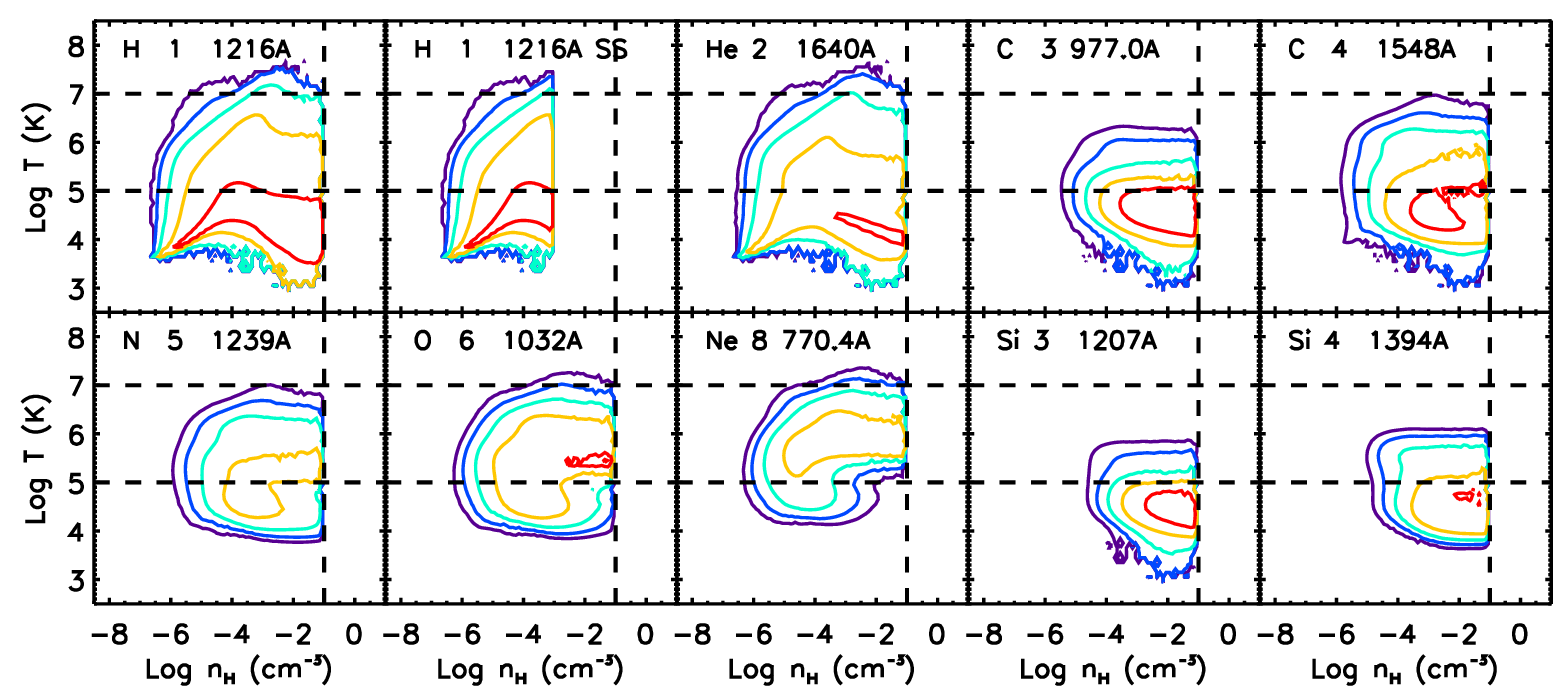}
\caption{The emission-weighted gas distributions in the temperature-density plane at $z=2$ for the \default\ simulation. From left to right, the panels show the distributions weighted by \hil, \hilSS, \heiih, \ciii\ and \civ\ emission (upper panels), and by \nv, \ovi, \neviii, \siiii\ and \siiv\ emission (lower panels). The horizontal and vertical lines are as in Fig.~\ref{twomass} and the contours are spaced logarithmically by 1.5 dex. Carbon and silicon lines trace mostly gas with $T<10^5\,$K, while \nv\ and \ovi\ better trace the WHIM with $10^5\,$K$<T<10^6\,$K. \neviii\ traces hotter gas, on average, than any other emission line in our sample, with $10^5\,$K $<T<10^7\,$K. The hydrogen and helium lines trace the gas mass better than the metal lines, although they are biased to cool, dense gas in close proximity to galaxies.}
\label{denstemp}
\end{figure*}

In this Section we investigate the physical properties of the gas that produces most of the emission. In particular, we want to understand what the typical densities and temperatures are of the gas that produces the bulk of the emission for each line. This will for example tell us whether the gas is bound to galaxies, resides in groups, or traces lower density regions like filaments and voids.

Fig.~\ref{twomass} shows the mass-weighted (left panel) and the metal mass-weighted (right panel) gas distributions in the temperature-density plane at $z=2$. Most of the mass resides in intergalactic gas with low density and temperature. Compared with the gas mass distribution, the distribution of metals in the IGM is biased towards higher densities and temperatures. As a consequence, the bulk of the metals do not trace the bulk of the mass.
Most of the metals are concentrated in two separate areas of the distribution: i) in the dense ($n_{\rm H}>0.1$ cm\3) star-forming gas, i.e.\ the ISM, for which we impose an effective equation of state and which we exclude from the analysis, and ii) in warm-hot gas in overdense regions with $10^{-6}$ cm\3 $< n_{\rm H} <10^{-3}$ cm\3.
This result is qualitatively similar to the findings of Paper I for $z=0.25$ (see also \citealt{wiersma2009b}).

Fig.~\ref{denstemp} shows the emission-weighted gas distributions in the temperature-density plane at $z=2$. From left to right, the panels show the distributions weighted by \hil, \hilSS, \heiih, \ciii\ and \civ\ (upper panels), \nv, \ovi, \neviii, \siiii\ and \siiv\  (bottom panels).

\begin{figure*}
\includegraphics[width=0.33\textwidth]{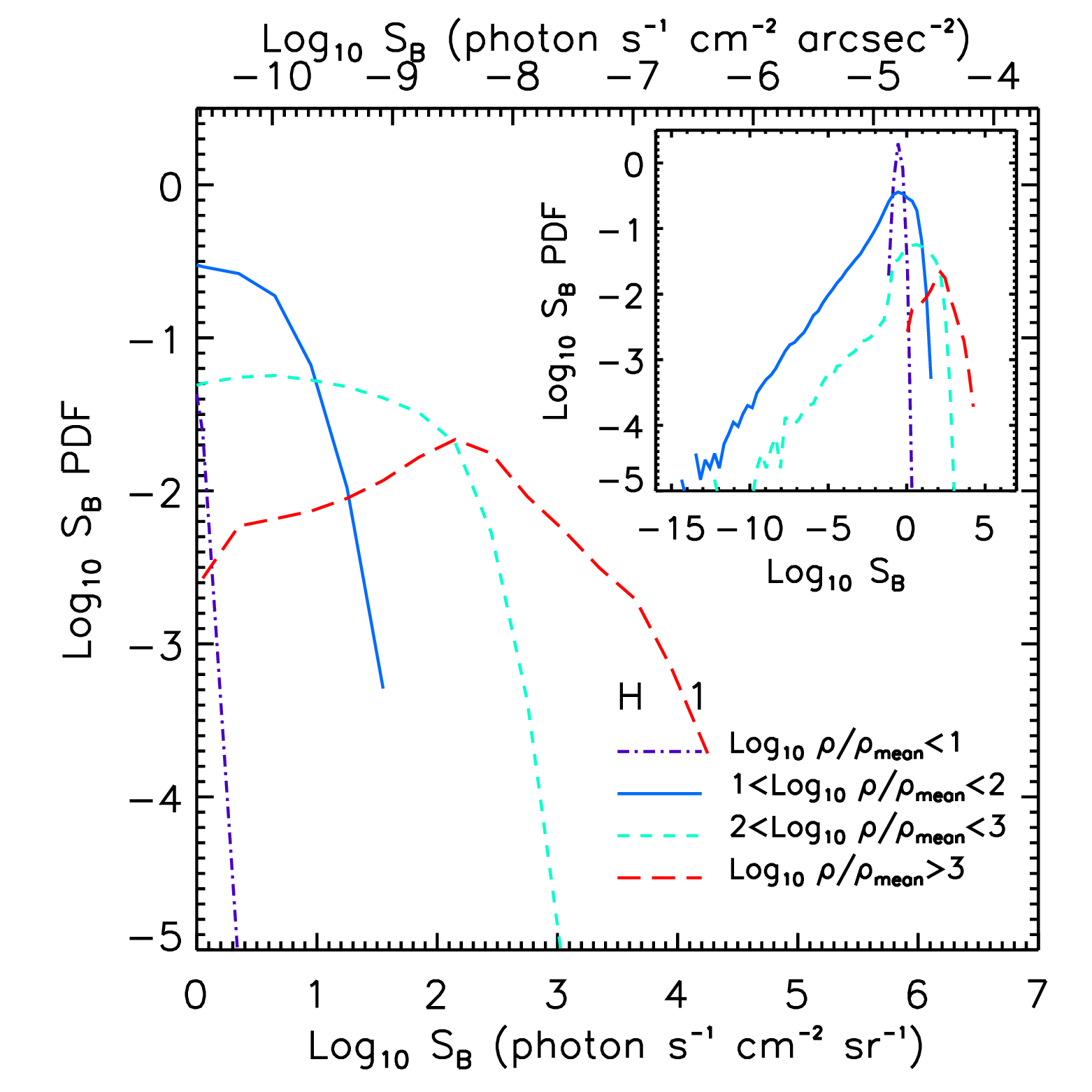} 
\includegraphics[width=0.33\textwidth]{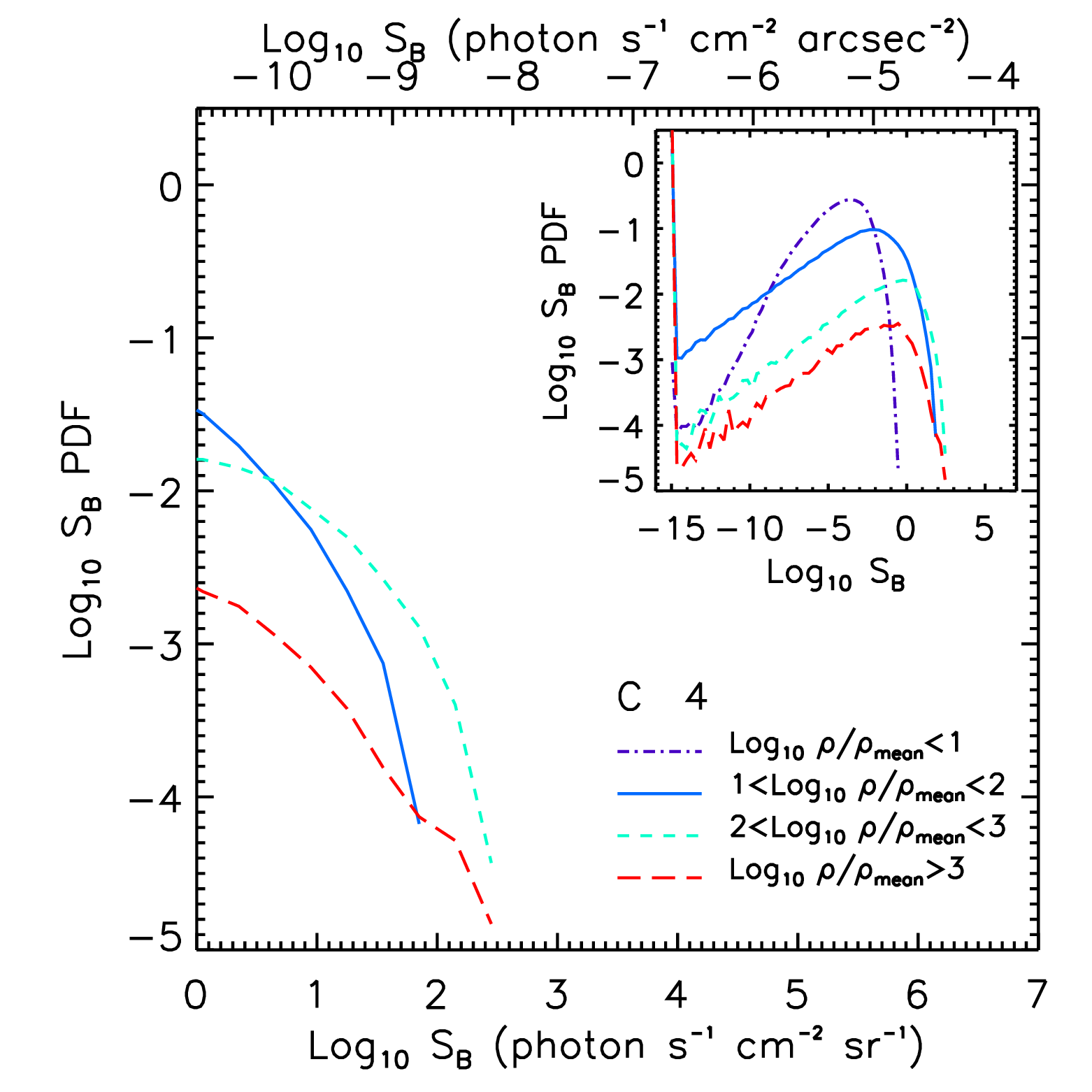} 
\includegraphics[width=0.33\textwidth]{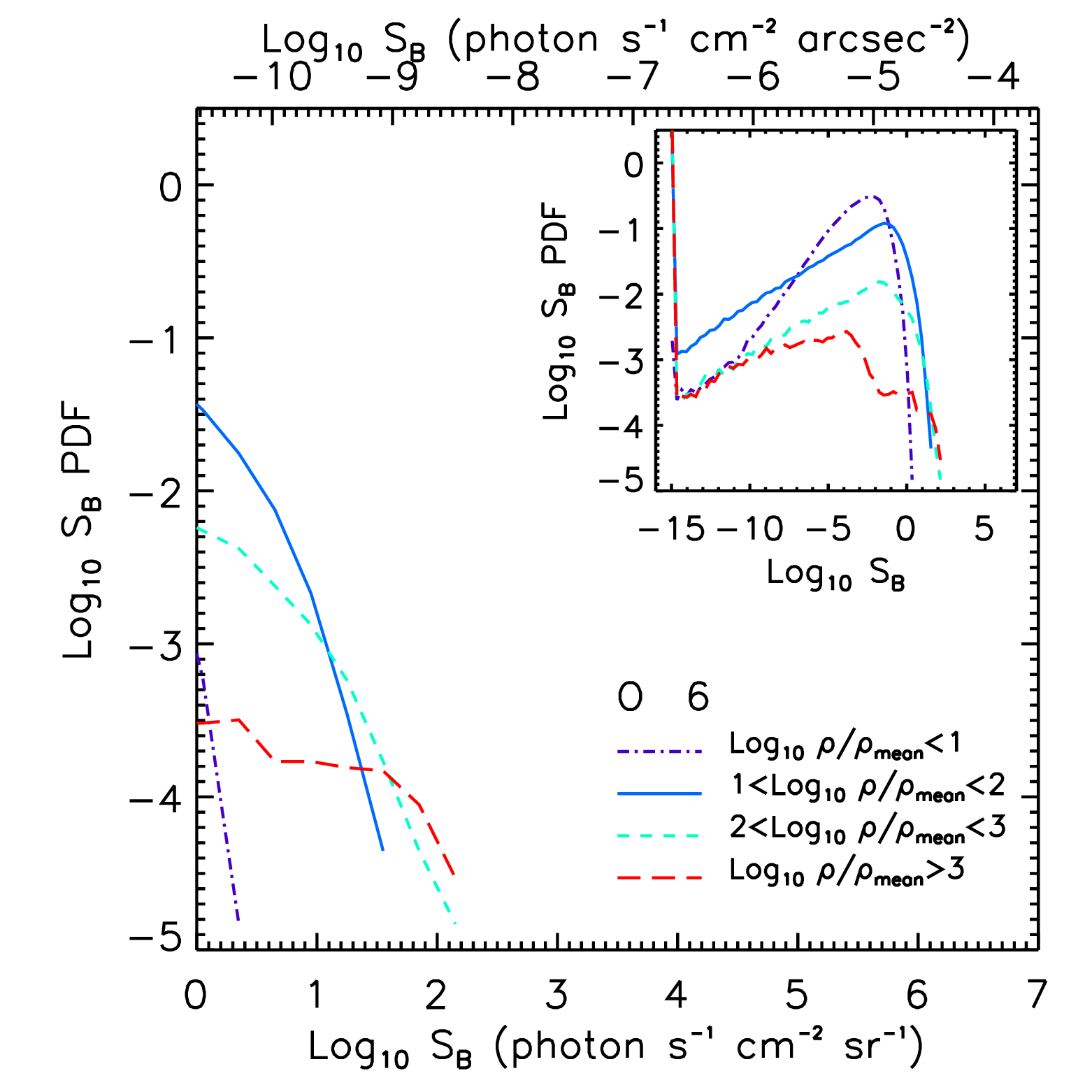}\\ 
\caption{As Fig.~\ref{lines_pdf}, but including only gas in fixed density intervals. From left to right the panels show the PDFs of the pixel surface brightness for \hil, \civ, and \ovi. While brightest \hil\ emission arises exclusively in highly overdense gas, a much wider range of overdensities contributes to the brightest \civ\ and \ovi\ pixels.}
\label{physcut}
\end{figure*}

\begin{figure*}
\includegraphics[width=0.33\textwidth]{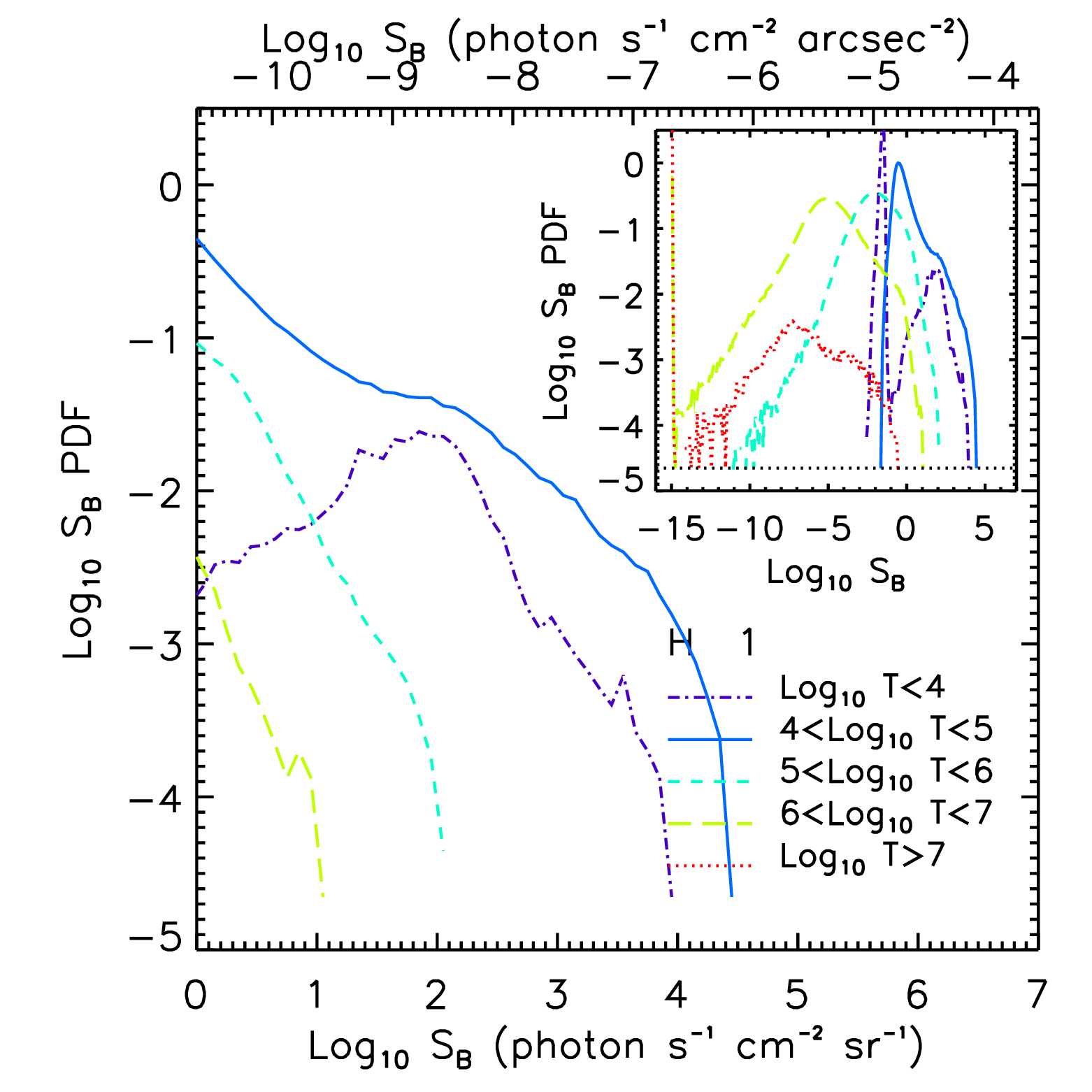} 
\includegraphics[width=0.33\textwidth]{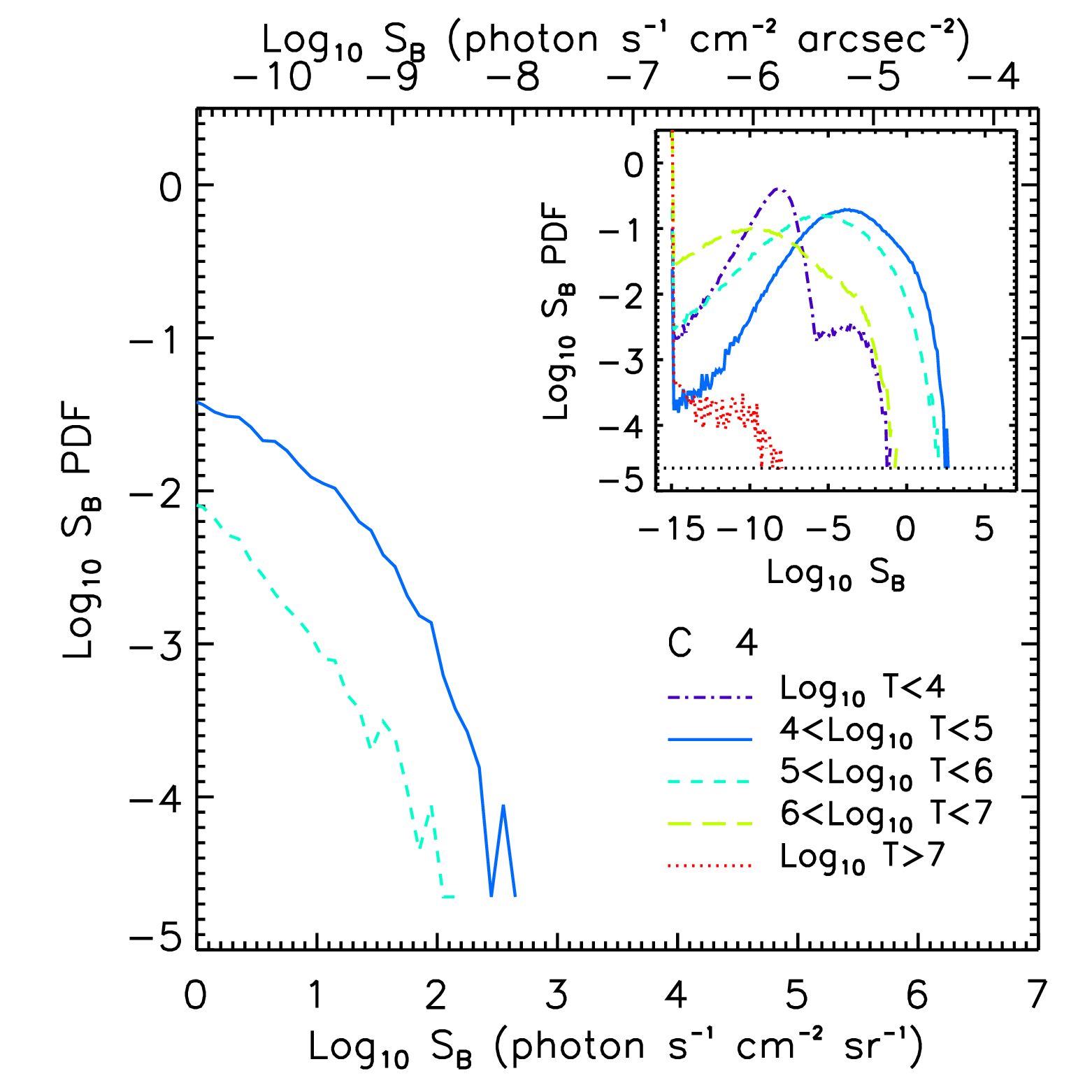} 
\includegraphics[width=0.33\textwidth]{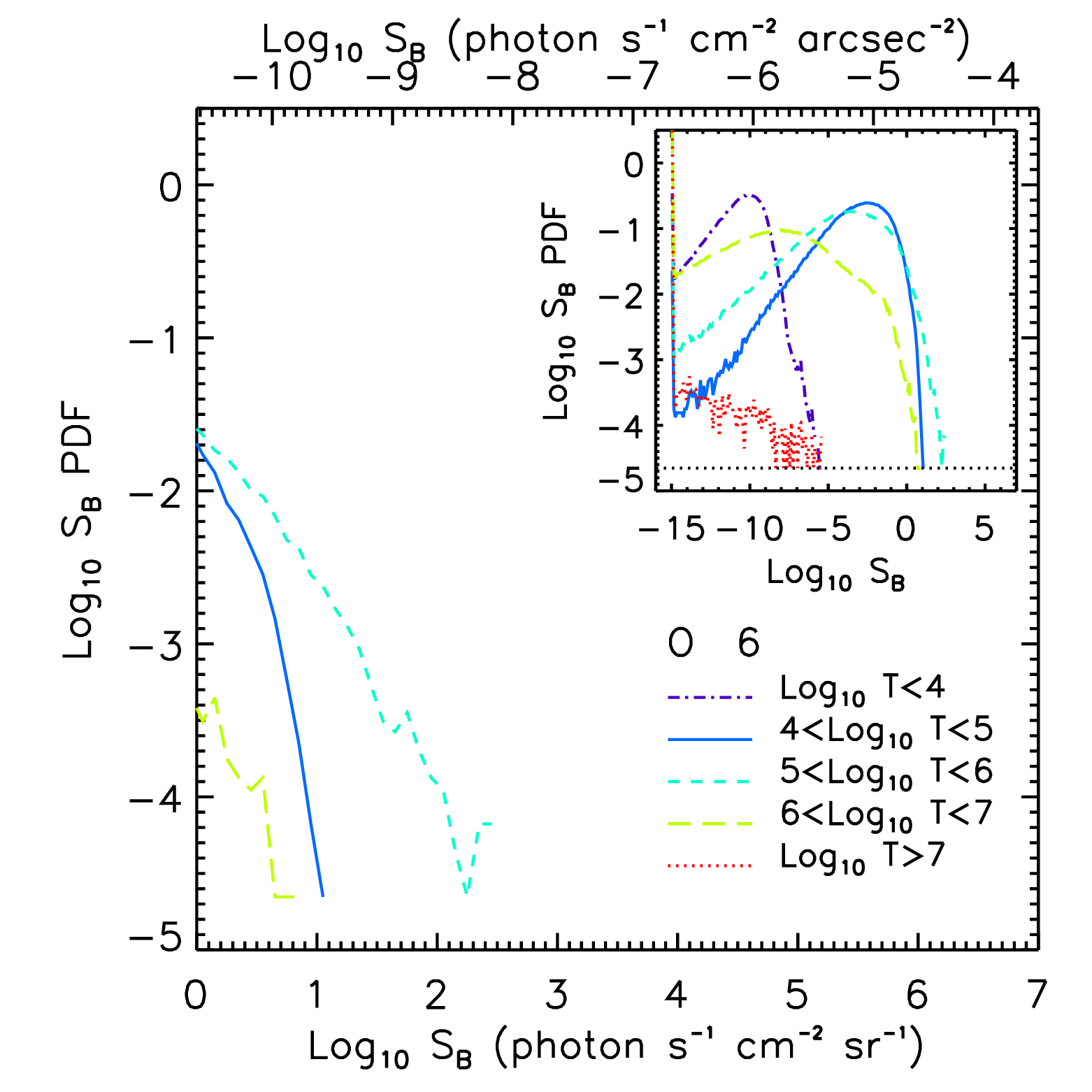} 
\caption{As Fig.~\ref{physcut}, but including only gas in fixed temperature intervals. For these metal lines the brightest emission comes from gas with temperatures close to the peak of the emissivity curves (see Fig.~\protect\ref{z2uv}).}
\label{physcut2}
\end{figure*}

A comparison of Figs.~\ref{twomass} and \ref{denstemp} demonstrates that the bulk of the emission from metal lines traces neither the IGM, nor the metal mass. This is consistent with what we found in Papers I and II at lower redshifts for X-ray and UV emission lines, respectively. Instead, each metal line traces a special region of the temperature-density plane centered on the temperature for which the line emissivity peaks (see Fig.~\ref{z2uv}). The width of the distribution in temperature corresponds roughly to the width of the emissivity curve as a function of temperature.
In terms of density and metallicity, all the emission-weighted distributions peak where the density is high, but also where both the metallicity and the density are high at the same time. The situation is different for the hydrogen and helium lines, which on average trace the gas mass well, once the quadratic dependence of the emissivity on the density is taken into account.

There are some notable differences between the different emission lines. Carbon and silicon lines trace gas with lower temperatures ($T<10^5\,$K) than other lines. \ovi\ traces diffuse gas with temperatures in the range $10^5\,$K $<T<10^6\,$K. \nv\ lines arise from gas with temperatures that fall between those of carbon and oxygen lines, while the \neviii\ lines trace hotter gas than any other line we consider here, with $T\sim 10^6\,$K.

Figs.~\ref{physcut} and \ref{physcut2} show explicitly how the surface brightness PDF varies with the gas density (Fig. \ref{physcut}) and temperature (Fig. \ref{physcut2}) for \hil\  (left columns), \civ\  (middle columns), and \ovi\  (right columns). We created these figures from emission maps that were computed by only including particles with the physical properties indicated in the legends. While \hil\ is highly biased towards high-density gas, with the highest surface brightness due to gas with overdensities $\rho/\left <\rho\right >\ga 10^3$, the brightest metal-line emission arises from gas with a wide range of overdensities, though still $> 10$. The gas responsible for the strongest \civ\ and \ovi\ emission has temperatures that are close to the peaks of the emissivity curves shown in Fig.~\ref{z2uv}, $T\sim 10^{5.0}\,$K and $\sim 10^{5.3}\,$K, respectively.

\section{Varying the physics}
\label{physics}

\begin{figure*}
\includegraphics[width=\textwidth]{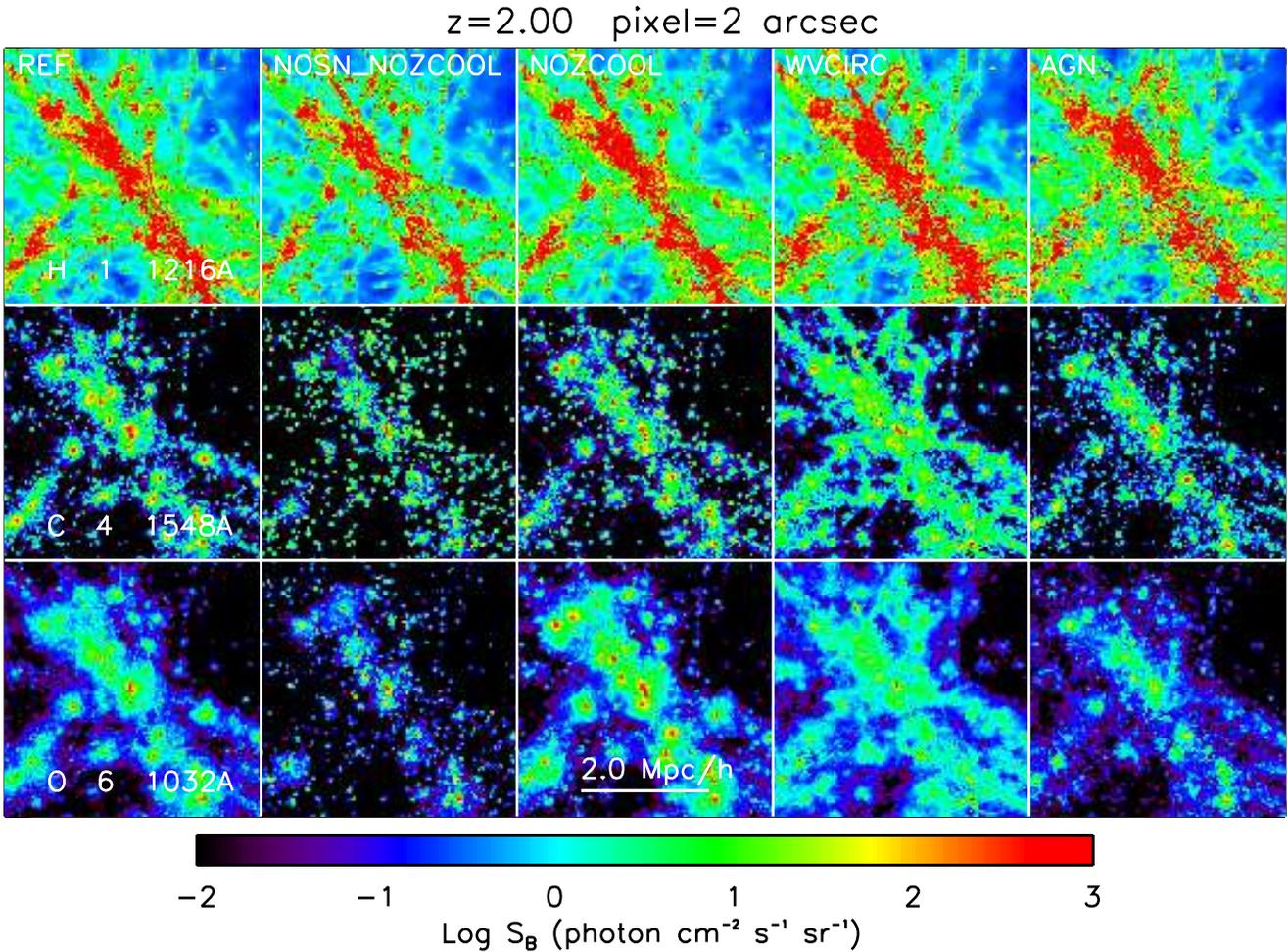}
\caption{As Fig.~\ref{allmaps}, but for a selection of \owls\ models with different physics implementations and for a selection of emission lines at $z=2$. From top to bottom, the lines shown are \hil, \civ\ and \ovi. From left to right, the results are for the \default, \nosn, \zcool, \wmom\ and \agn\ models respectively.}
\label{four1}
\end{figure*}

\begin{figure*}
\includegraphics[width=\textwidth]{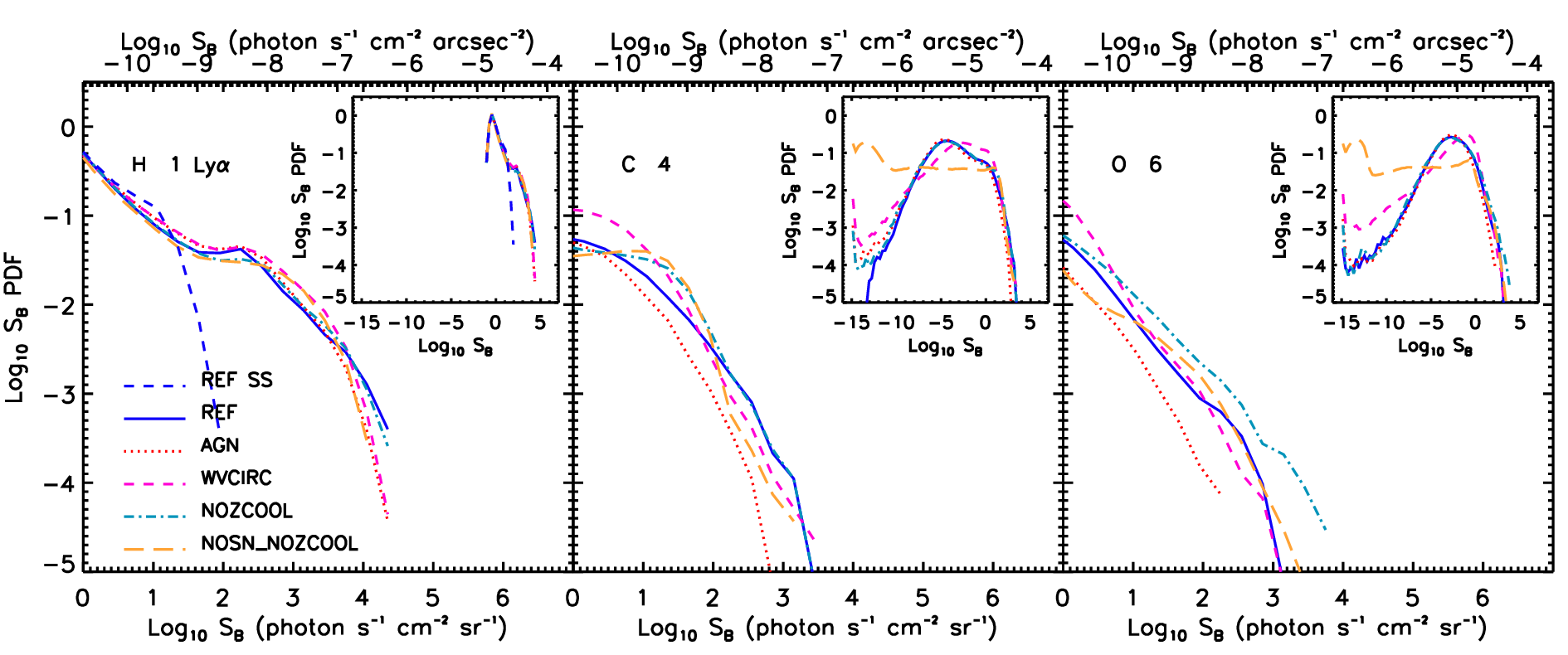}
\caption{As Fig.~\ref{lines_pdf}, but for a selection of \owls\ models that include different physical processes. From left to right, the different panels are for $z=2$ emission from \hil, \civ\ and \ovi, respectively. There is almost no variation in the \hil\ emission predicted by the different simulations, although removing emission from gas with $n_{\rm H} > 10^{-3}\,{\rm cm}^{-3}$, as would be appropriate if no \lya\ photons would be produced in or escape from such regions, would cut off the bight end of the distribution. 
The differences are largest for low surface brightness metal-line emission, but the models tend to converge at the bright end. Metal-line cooling is particularly important for \ovi. AGN feedback slightly decreases the peak surface brightness of both \civ\ and \ovi.}
\label{both_pdf}
\end{figure*}

In this Section we discuss the influence of different physical processes on the results. We focus our analysis at $z=2$, but all our conclusions are valid for results at higher redshifts. We consider five different sets of \owls\ runs, in which the physical modules have been changed one at a time. The models are as follows (see \citealt{schaye2010} for more detailed descriptions):

\begin{enumerate}
\item \default: Our reference model, as described in Section~\ref{owls}. All results presented in the preceding sections were obtained from this simulation. This model uses the star formation prescription  of \citet{schaye2008}, the supernova-driven wind model of \citet{vecchia2008} with initial wind mass loading $\eta = 2$ and initial wind velocity $v_{\rm w}=600$ km s\1, the metal-dependent cooling rates of \citet{wiersma2009a}, the chemical evolution model described in \citet{wiersma2009b} and the WMAP3 cosmology. AGN feedback is not included.

\item \nosn: As \default, but without supernova feedback and with radiative cooling rates that assume primordial abundances. The metals produced by stars are thus only transported by gas mixing and no energy is transferred to the IGM and ISM when SNe explode.

\item \zcool: As \default, but with radiative cooling rates calculated assuming primordial abundances. Neglecting the contributions of metal lines implies slower gas cooling than in the \default\ model.

\item \wmom: As \default, but with ``momentum--driven'' galactic winds following the prescription of \citet{oppenheimer2008}, with the difference that, as in the \default\ model, winds are ``local'' to the star formation event and are fully hydrodynamically coupled (see \citealt{vecchia2008} for a discussion of the importance of these effects). The wind velocity and mass loading factor are functions of the galaxy velocity dispersion $\sigma$: $v_{\rm w} = 5\sigma$ and $\eta = v_{\textrm{w}0}/\sigma$, with $v_{\textrm{w}0}=150$ km s\1. The velocity dispersion $\sigma = \sqrt{2}v_{\rm circ}$ is estimated using an on-the-fly friends-of-friends halo finder. This model was motivated by the idea that galactic winds may be driven by radiation pressure on dust grains rather than SNe. The energy injected into the wind becomes much higher than for \default\ for halo masses exceeding $10^{11} - 10^{12}\,{\rm M}_\odot$.

\item \agn: As \default, but with the addition of AGN feedback, using the prescription of \citet{booth2009}.
\end{enumerate}

We chose this set of simulations because these are the models that produced the largest variations in Papers I and II.
In these papers, which focused on $z=0.25$, we found that for most models the variations in the results are small relative to the \default\ model, especially for UV lines. In terms of the bright-end tail of the surface brightness PDF, the largest discrepancies with the \default\ results were found for model \zcool. The inclusion of AGN feedback could decrease the X-ray emission at the centres of groups and clusters by up to a factor of $\sim 100$, but barely affected the emission from diffuse IGM gas. Paper II also showed that varying the feedback parameters that control the intensity of the winds (e.g.\ the wind mass loss rate and velocity) introduced variations in the UV results of a factor of a few or less.

Fig.~\ref{four1} shows maps of the surface brightness of \hil, \civ\ and \ovi\ lines in the different simulations. The PDFs of the surface brightness are shown in Fig.~\ref{both_pdf}. All maps assume an angular resolution of 2" and show results at $z=2$. Most models predict comparable maximum surface brightness values, as can be seen by comparing the high surface brightness tails of the PDFs ($S_{\rm B} > 100$ \phot), but the variations can be very large for fainter metal-line emission. For \hil, on the other hand, the differences are always very small.

In the \nosn\ model the emission is concentrated close to galaxy haloes, while the surface brightness of low-density regions is extremely low for metal lines. This produces the flattening observed in the PDFs of metal lines for $S_{\rm B} < 0-100$ \phot\ and is due to the inefficient transport of metals from galaxies to the diffuse IGM in the absence of SN feedback.

Model \zcool\ predicts somewhat stronger emission from metals than the \default\ model, but comparable \hil\ emission. In dense regions, the \ovi\ emission exceeds the predictions of the \default\ model by up to an order of magnitude. This indicates that metal-line cooling is important in the gas for which the \ovi\ emission peaks, i.e.\ highly enriched gas with $T\sim 10^{5.3}\,$K. Note that ignoring metal-line cooling is not self-consistent, as it allows the gas to linger too long at the temperatures for which the emissivity is high.

For model \wmom\ the peaks of the metal-line PDFs are shifted to higher surface brightness values. In other words, compared with model \default, a larger fraction of the sky, and a larger fraction of the mass in the Universe, shines bright in the UV. The outflows in model \wmom\ are more efficient at transporting metals into the IGM, mainly as a result of the higher mass loading in low-mass galaxies. As such, the different spatial distribution of the UV emission we see in Fig.~\ref{four1} between the \default\ and \wmom\ models is mostly due to differences in the metallicity of the gas, and to a lesser extent to differences in the temperature and density. In principle, the detection of diffuse metal-line emission could thus help decide which one of the descriptions of the physics of galactic winds is more ``realistic''. A larger spatial extent and a more filamentary structure of the emission on large scales would weigh in favour of highly mass-loaded galactic winds.

Finally, the predictions of model \agn\ are comparable to those of the \default\ model in regions with low surface brightness and low density. However, at the bright tail of the distribution model \agn\ predicts surface brightnesses that are up to 0.5~dex lower. This is a consequence of the fact that AGN feedback decreases the density and metallicity of the gas near massive galaxies.

The relatively small differences between the metal-line emission predicted by models that include very different physical processes gives us confidence that our predictions for the brighter emission are robust to factors of a few. On the other hand, the fact that the differences in the predictions for the diffuse metal-line emission are not negligible, implies that observations of this faint glow have the potential to discriminate between different models for feedback from star formation and AGN. 

Finally, we recall that while the different simulations predict very similar \hil\ emission, the predictions for this line are uncertain due to the importance of self-shielding from ionising radiation, resonant scattering, and absorption by dust, which are all ignored here. The potential impact is illustrated by the blue dashed curve labelled `REF SS' in the left panel of Fig.~\ref{both_pdf}, which indicates the surface brightness PDF if we ignore all emission from gas with $n_{\rm H}> 10^{-3} \,{\rm cm}^{-3}$, an assumption that is very conservative for our purposes.

\section{Conclusions}
\label{conclusion}

We have investigated the nature and detectability of redshifted rest-frame UV line emission from the IGM (i.e.\ $n_{\rm H}<10^{-1}\,{\rm cm}^{-3}$) at $2<z<5$. While more strongly biased towards high-density gas than absorption lines, emission lines offer the possibility to directly map the distribution of the gas and metals in three dimensions.
Following the procedures of \citet{bertone2010a}, we used a set of large, cosmological, hydrodynamical simulations taken from the \owls\ project \citep{schaye2010} to create simulated maps of a set of strong rest-frame UV emission lines. We focused on \ciii, \civ, \nv, \ovi, \neviii, \siiii\ and \siiv\ lines (see Table~\ref{eltable}), which are the strongest metal lines that trace the diffuse IGM and that can potentially be detected by current and upcoming optical instruments. For reference, we also showed results for \hil\ and \heiih\ and for the former we illustrated the potential impact of self-shielding and dust extinction by considering only the flux emitted by gas with $n_{\rm H}<10^{-3}\,{\rm cm}^{-3}$, which provides a conservative estimate of the minimum signal that can be expected.  

Our main results can be summarised as follows:
\begin{enumerate}
\item Depending on the line and the redshift, we predict a maximum surface brightnesses of order $10^1 - 10^5$ \phot\ in the redshift interval $2\le z \le 5$ for 2'' pixels (see Fig.~\ref{redshift_pdf}).  The strongest IGM emission comes from \hil, provided we include emission from gas with $10^{-3}\,{\rm cm}^{-3} < n_{\rm H} < 10^{-1}\,{\rm cm}^{-1}$. The highest surface brightness metal-line emission comes from \ciii\ and is about an order of magnitude fainter. The highest surface brightness regions for \civ, \siiii, \siiv\ and \ovi\ lines are fainter than the brightest \ciii\ by factors of a few. The \nv\ and \neviii\ lines, as well as \heiih, are substantially fainter but their maximum surface brightnesses still exceed $10^2$ \phot\ for $z=2$ (and 2'' pixels).

\item Lines from different ions trace gas in different temperature regimes. The gas responsible for the highest surface brightness
metal-line emission typically has temperatures close to the peak of
the corresponding emissivity curve. \hil, \heiil, \ciii, \siiii\ and \siiv\ lines prevalently trace gas with $T < 10^5\,$K near galaxies, making them excellent probes of cold accretion flows and the colder parts of galactic winds. \civ\  ($T\sim 10^5\,$K), \nv\  ($T\sim 10^{5.1}\,$K), \ovi\  ($T\sim 10^{5.4}\,$K), and \neviii\  ($T\sim 10^{5.8}\,$K) trace the warmer gas in haloes and filaments and thus constitute powerful tools to study the WHIM and the warmer parts of outflows. 

\item While the brightest \hil\ emission arises exclusively in highly overdense gas ($\rho/\left <\rho\right >\ga 10^3$), at least if we assume the gas to be optically thin, a much wider range of overdensities contributes to the brightest pixels of \heiil\ and high-ionisation metal lines such as \civ, \ovi\ and \neviii. 
The spatial distribution of the emission is related to the ionisation state of the emitting gas. Lower ionisation gas produces clumpier and more clustered emission, while higher ionisation gas produces smoother and more diffuse emission profiles. The degree of ionisation decreases with increasing density but increases with the temperature. As the emissivity increases with the gas density, emission from lower-ionisation gas tends to be brighter. Indeed, we find for example that the beryllium-like atoms \niv\ and \ov\  (which were not shown here because they only shift into the optical waveband for very high redshifts) are stronger than their higher-ionisation lithium-like counterparts. 

\item We have compared our fiducial simulation with models that ignore SN feedback and/or metal-line cooling, with a model that includes highly mass-loaded ``momentum-driven'' galactic winds, and with a simulation that takes feedback from AGN into account. While the faint metal-line emission is very sensitive to the inclusion and implementation of galactic winds (because these are responsible for enriching the low-density IGM with metals, where the weak emission is produced), the results for bright metal-line emission, and even more so for both faint and bright \hil\ emission, are remarkably robust to changes in these physical processes. Accurate cooling rates are required to predict the bright metal-line emission with a precision better than a factor of a few, particularly for lines that peak at temperatures for which the cooling time is short, such as \ovi. AGN feedback primarily affects the emission from the cores of haloes, reducing the brightest metal-line emission by up to factors of a few. Highly mass-loaded galactic winds, on the other hand, can boost the brightest metal-line emission by similar factors. 

\item According to our predictions, \muse\ and \kcwi\ should be able to detect \hil\ emission from regions with $\rho > 10^3 \bar{\rho}_{\rm b}$ up to at least $z=5$ (provided our optically thin calculation is appropriate in this regime), while \hil\ from less dense regions and the \heiih\ line should be visible at lower redshifts. \civ, \ovi\ and \siiv\ will all be detectable up to $z\approx 3$, while \ciii\ and \siiii\ should be detectable up to $z\approx 4$. The \nv\ and \neviii\ lines should also be within the capabilities of these instruments. We note, however, that lines with rest-frame wavelengths shorter than that of \lya\ (i.e.\ \ciii, \ovi, \neviii, and \siiii) may suffer substantial foreground absorption, which we have not accounted for here. \ov\ lines at $z\approx 2.2$ and \ciii\ lines at $z\approx 1$ are roughly within the sensitivity of the \fireball\ experiment, and should be considered as possible contaminants of \lya\ emission.

We conclude that several rest-frame UV emission lines from the high-redshift IGM will become detectable in the near future, possibly starting with the \cwi\ which is already operating on Palomar. Observations of these lines have the potential to revolutionise studies of the interactions between high-redshift galaxies and their environment. Lower ionisation lines provide us with tools to image cold accretion flows as well as cold, outflowing clouds. Higher ionisation lines, on the other hand, will allow us to image the metal-enriched WHIM and galactic winds. Such observations, particularly the simultaneous detection of multiple lines, will provide strong constraints on the density, temperature, and chemical composition of the gas. 

\end{enumerate}

\section*{Acknowledgments}
We are grateful to all members of the \owls\ team for help during the development of this project and to Tom Theuns and Freeke van de Voort also for a careful reading of the manuscript. We would like to thank Xavier Prochaska and Christopher Martin for stimulating discussions and we are grateful to the anonymous referee for a helpful report. The simulations presented here were run on Stella, the LOFAR BlueGene/L system in Groningen, on the Cosmology Machine at the Institute for Computational Cosmology in Durham as part of the Virgo Consortium research programme, and on Darwin in Cambridge. This work was sponsored by the Dutch National Computing Facilities Foundation (NCF) for the use of supercomputer facilities, with financial support from the Netherlands Organization for Scientific Research (NWO). SB acknowledges support by NSF Grants AST-0507117 and AST-0908910. This work was supported by an NWO VIDI grant and by the Marie Curie Initial
Training Network CosmoComp (PITN-GA-2009-238356).

\appendix

\section{Angular resolution}
\label{angle}

\begin{figure*}
\centering
\includegraphics[width=0.33\textwidth]{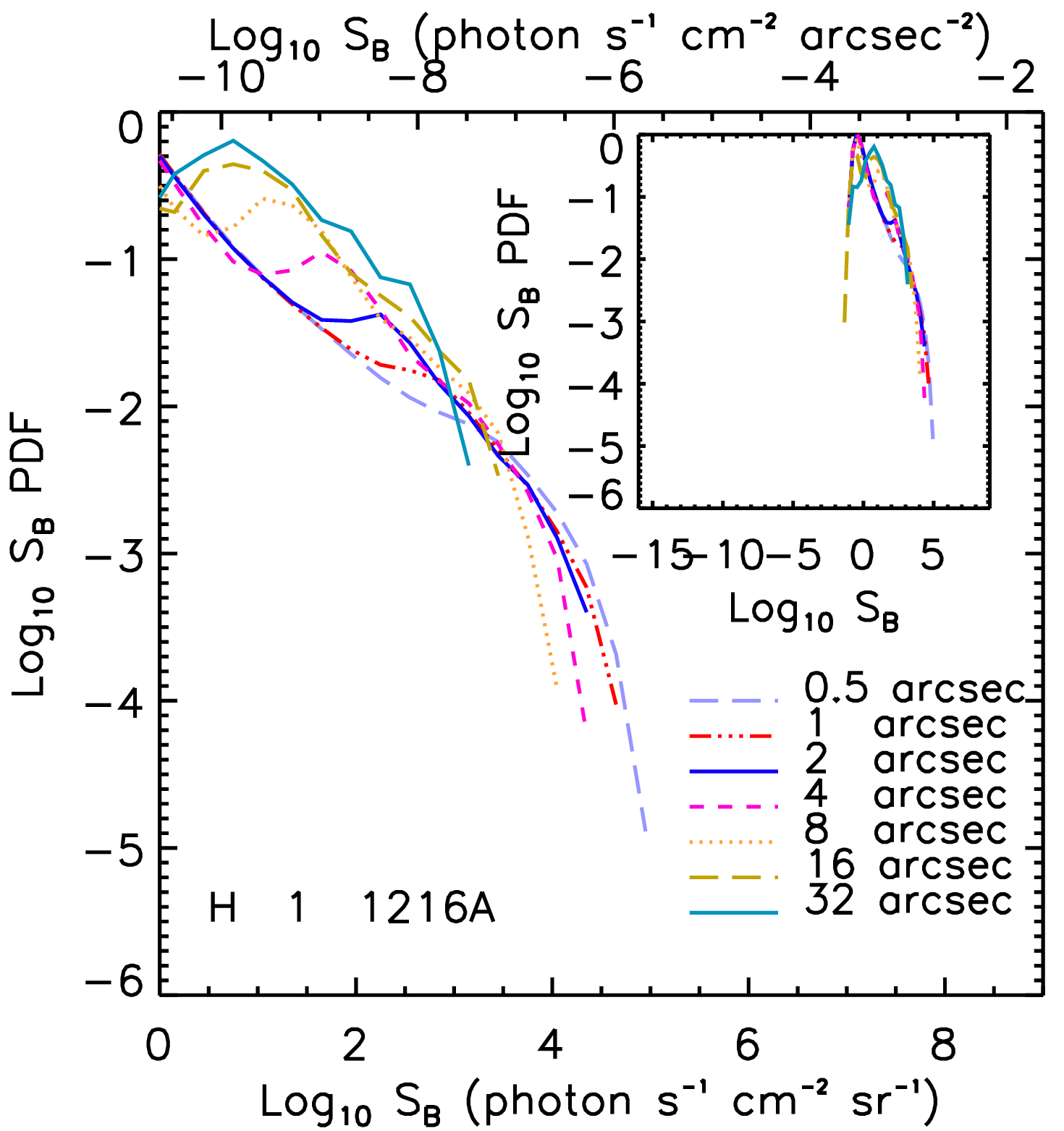}
\includegraphics[width=0.33\textwidth]{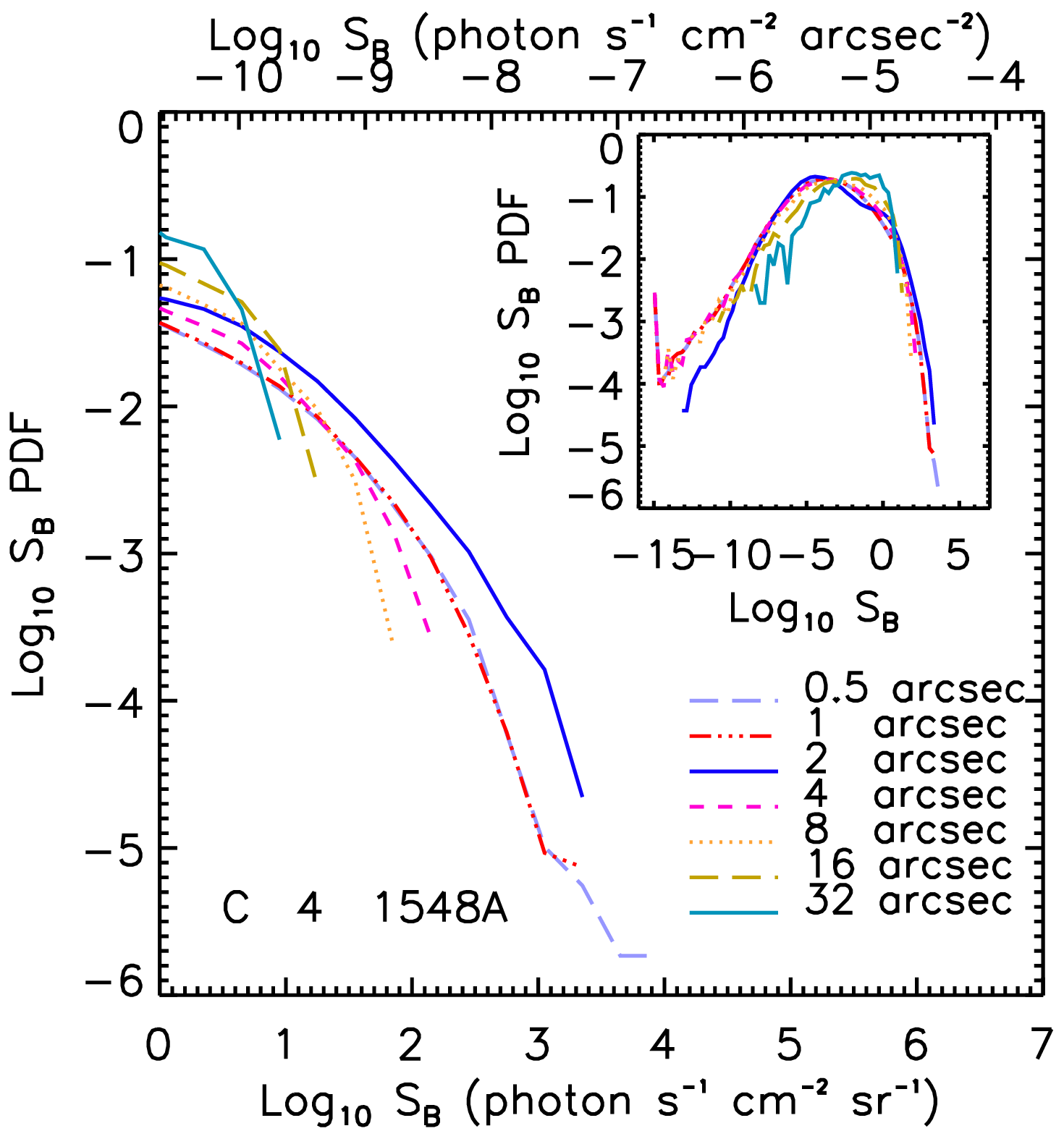}
\includegraphics[width=0.33\textwidth]{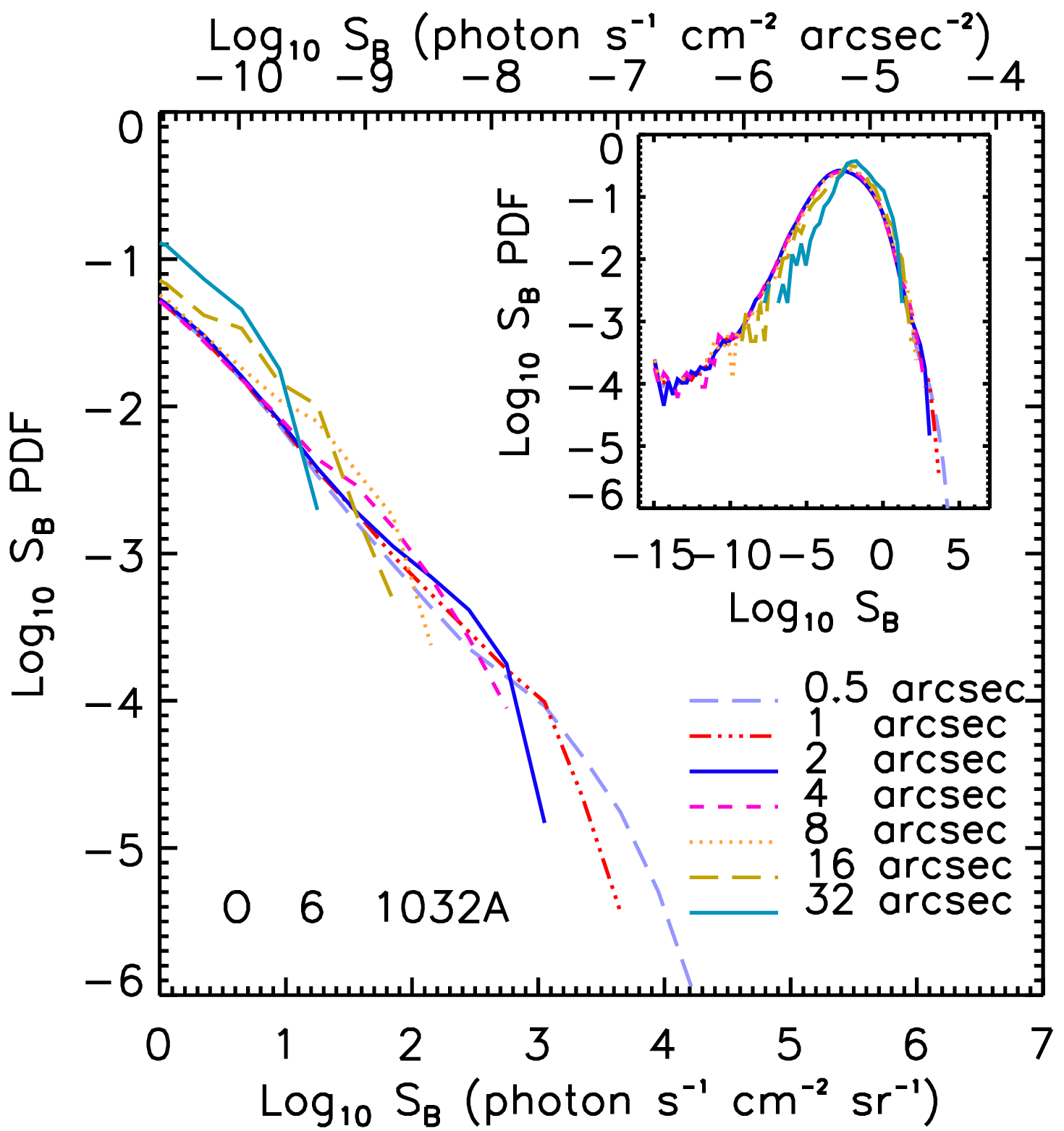}
\caption{As Fig.~\ref{lines_pdf}, but for \hil\ (left panel), \civ\ (middle panel), and \ovi\ (right panel) emission as a function of the angular resolution of the emission maps. The pixel sizes are $\vartheta = 0.5"$, $1"$, $2"$, $4"$, $8"$, $16"$, and $32"$, corresponding to physical sizes of 3.1, 6.2, 12.4, 24.8, 49.7, 99.3, and 199 \hm\ kpc, respectively, at $z=2$. 
The maximum predicted surface brightness increases steadily with the angular resolution. The results are less sensitive to the pixel size for lines that arise in lower density gas, such as \ovi.}
\label{angle_figure}
\end{figure*}

In this Appendix we test the effect of varying the angular resolution used to make the emission maps. Fig.~\ref{angle_figure} compares the PDFs of the surface brightness of, from left to right, \hil, \civ, and \ovi\ at $z=2$ for maps with resolutions $\vartheta = 0.5", 1", 2", 4", 8", 16"$, and $32"$, which correspond to physical scales varying from 3.1 \hm\ kpc to 199 \hm\ kpc.

The distributions converge well for all but the highest surface brightness values. The increase in the maximum brightness with decreasing pixel size reflects the fact that large pixel sizes smooth the strong emission coming from structures with an angular size smaller than that used to observe them. The differences between the PDFs corresponding to different pixel sizes are much smaller for \ovi\ than for \civ\ and \hi. This is because \ovi\ emission arises in gas of lower densities, which will have larger coherence lengths (because the Jeans scale increases with decreasing density; see e.g.\ \citealt{schaye2001}).

\section{Convergence tests}
\label{converge}

In this Appendix we present a number of tests aimed to verify how resolution affects the numerical predictions. We consider the effect of varying the numerical resolution in Appendix~\ref{number}, the box size in Appendix~\ref{box} and the slice thickness in Appendix~\ref{thick}. All tests use the \default\ model and results are shown only for the \ovi\ 1032 \AA\ line.

\subsection{Numerical resolution}
\label{number}

\begin{figure*}
\centering
\includegraphics[width=0.33\textwidth]{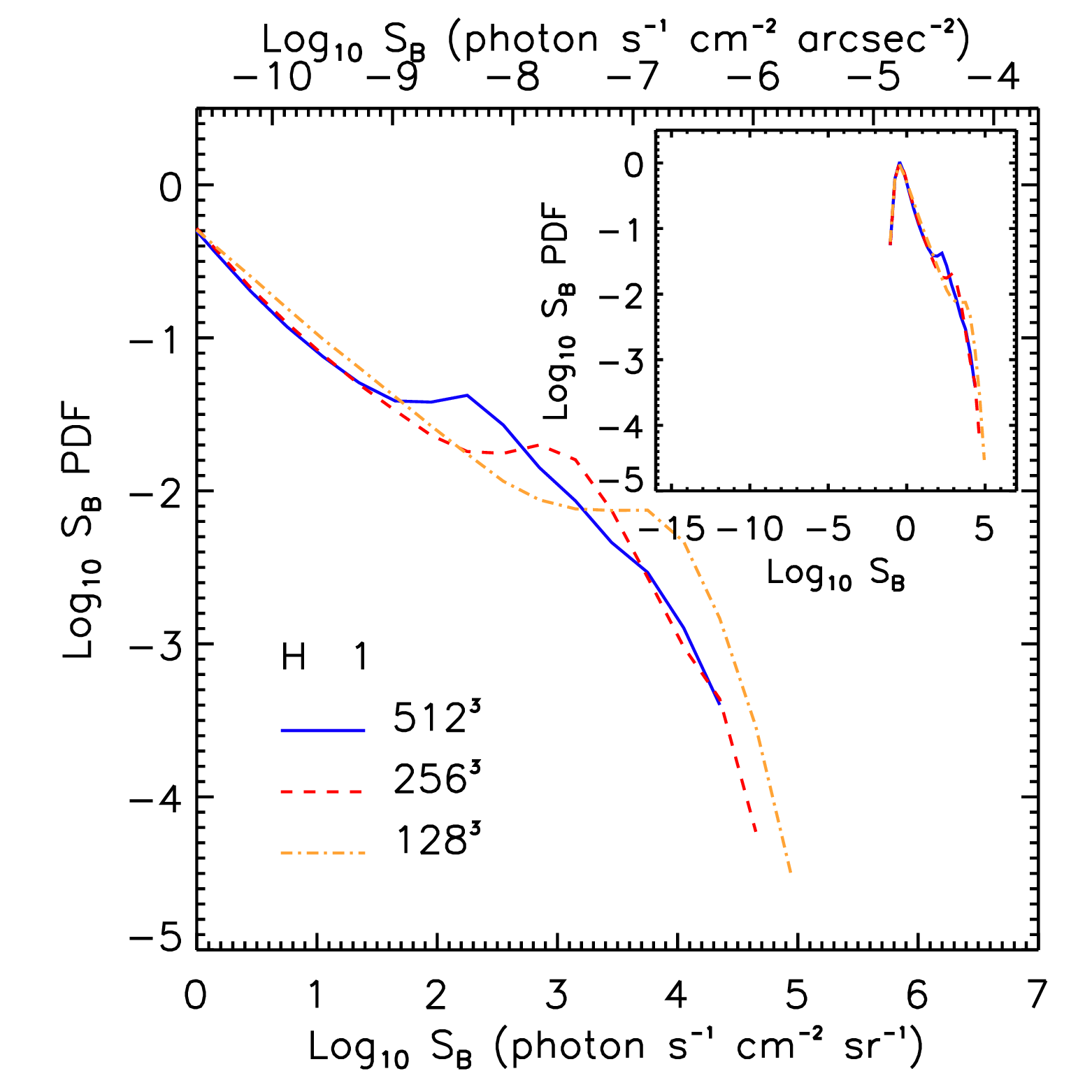}
\includegraphics[width=0.33\textwidth]{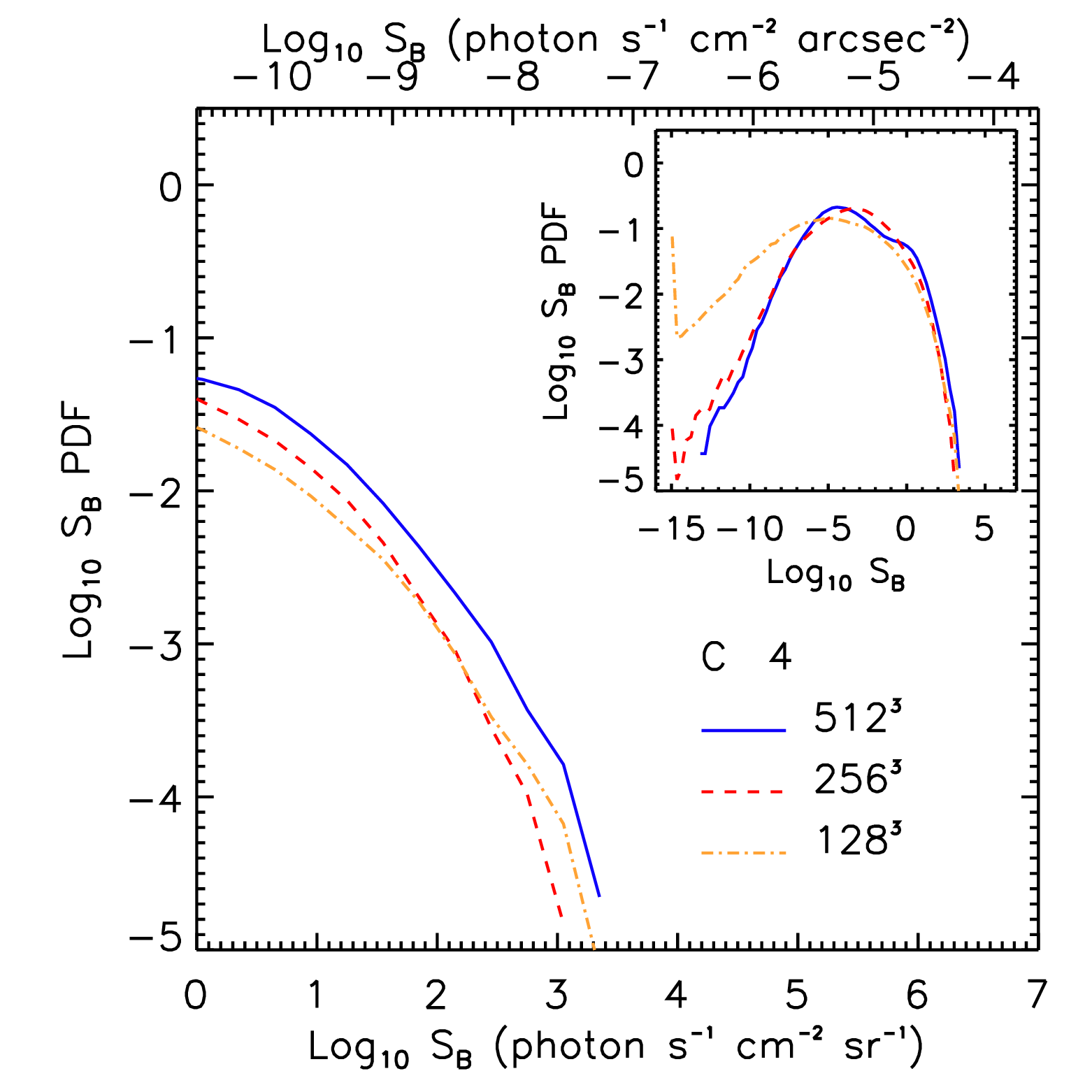}
\includegraphics[width=0.33\textwidth]{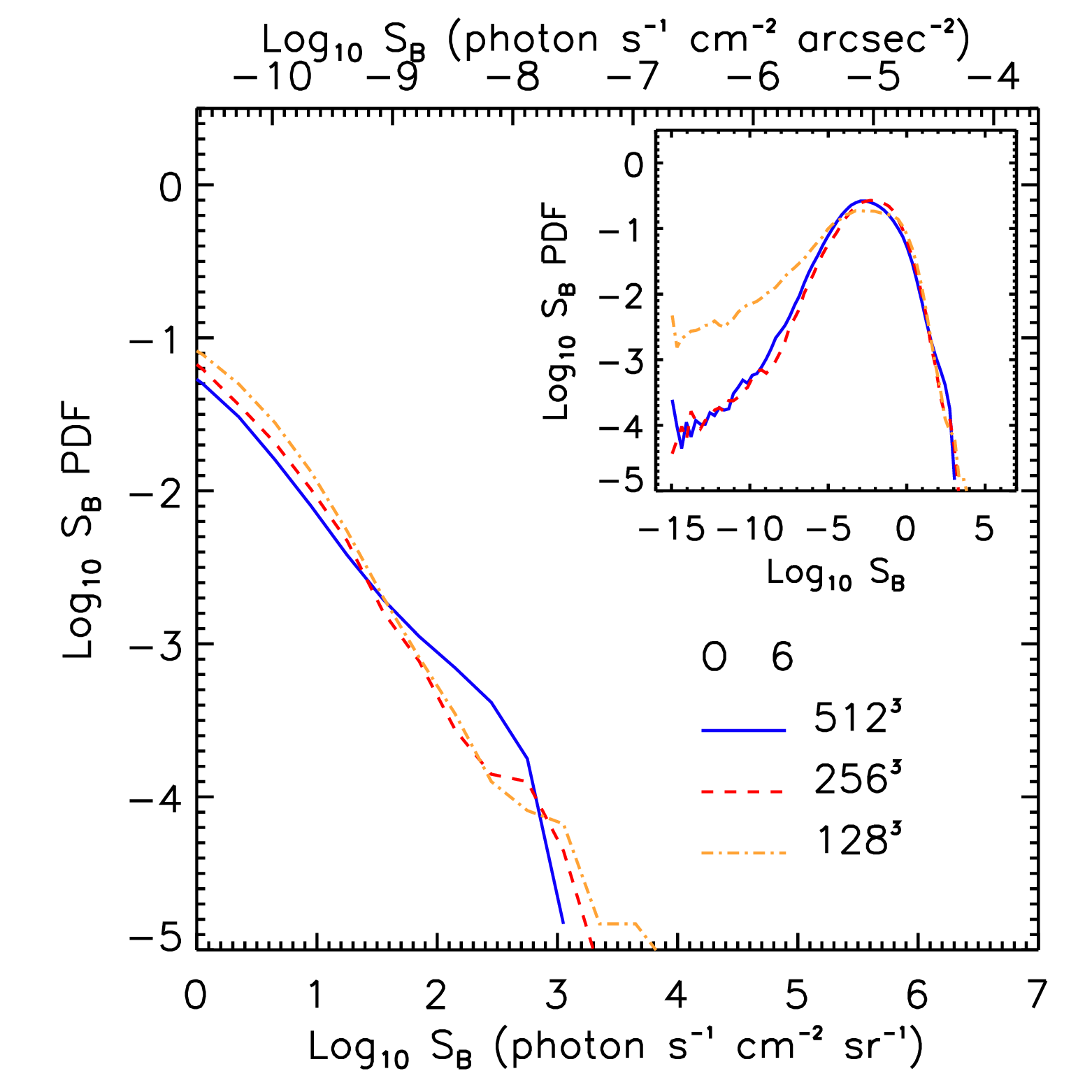}
\caption{As Fig.~\ref{lines_pdf}, but for \hil\ (left panel), \civ\ (middle panel), and \ovi\ (right panel) emission as a function of the numerical resolution of the simulation. The particle mass increases by a factor of eight and the force resolution decreases by a factor of two when the number of particles decreases by a factor of $2^3$. Our fiducial run is converged with respect to the numerical resolution.}
\label{number_figure}
\end{figure*}

To test the effect of varying the numerical resolution, we compare emission maps for three different realisations of the same simulation with particle numbers $N=2\times 128^3$, $2\times 256^3$ and $2\times 512^3$. The corresponding dark matter particle masses are $m_{\rm DM} = 4.1\times 10^{8}$, $5.1\times 10^7$, and $6.3\times 10^6$ \hm\ \msun, respectively.

Fig.~\ref{number_figure} shows the PDFs of the $z=2$ surface brightness for, from left to right, \hil, \civ, and \ovi\  for maps with 2" angular resolution. All distributions are similar at high and intermediate surface brightnesses. Only the lowest resolution run with $N=2\times 128^3$ deviates strongly from the other two and only for $<10^{-5}$ \phot. We conclude that our conclusions are not significantly affected by the finite resolution of our simulations.

\subsection{Box size}
\label{box}

\begin{figure*}
\centering
\includegraphics[width=0.33\textwidth]{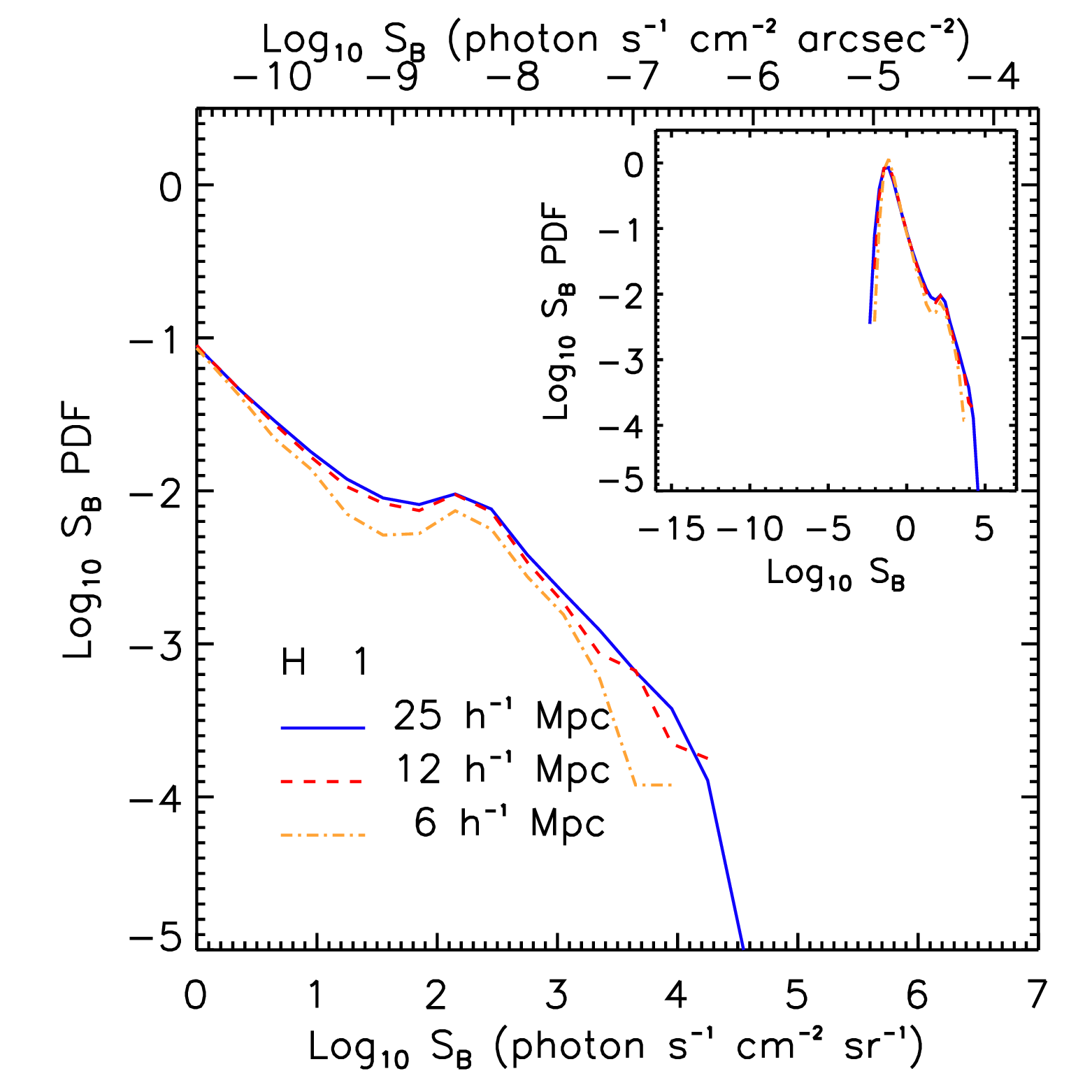}
\includegraphics[width=0.33\textwidth]{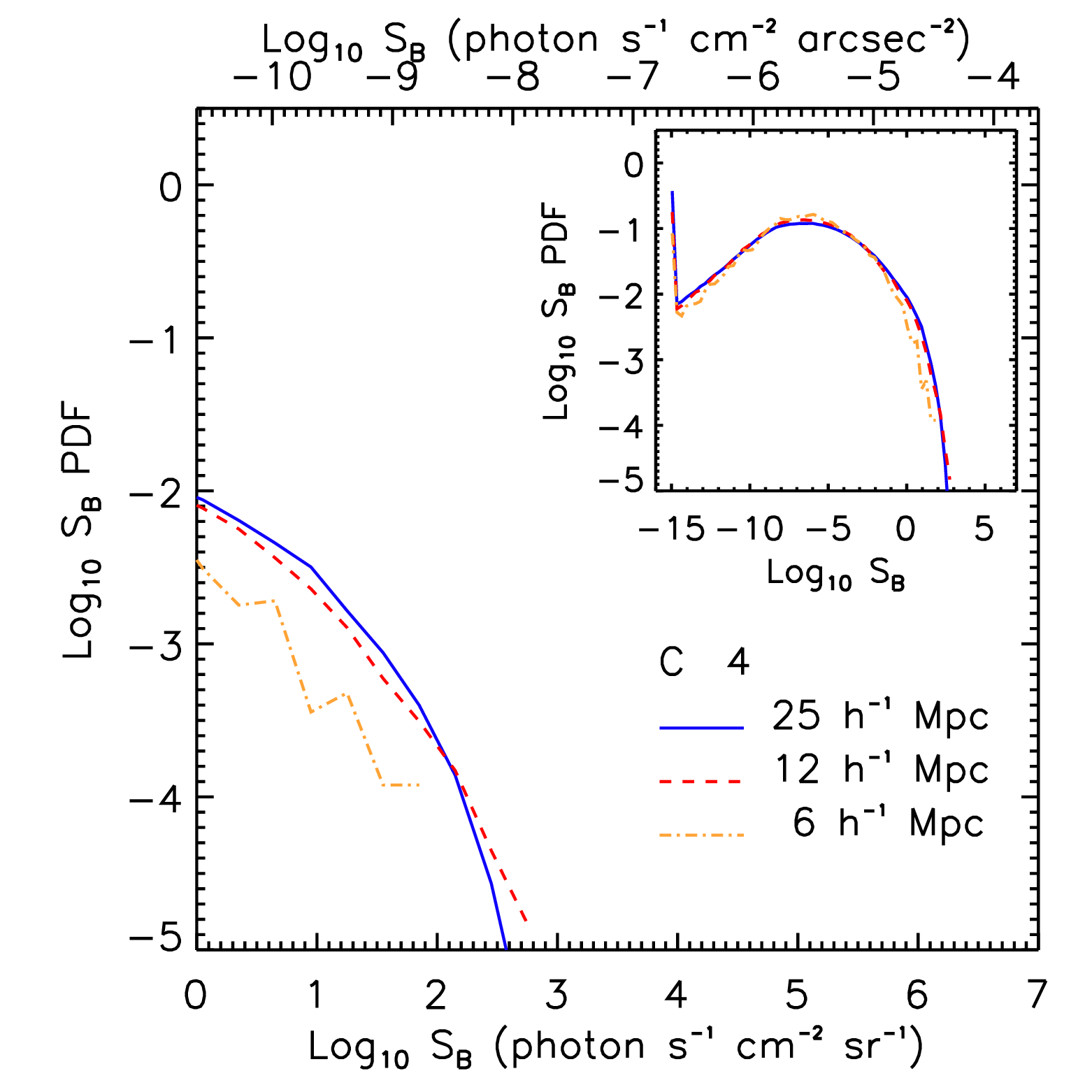}
\includegraphics[width=0.33\textwidth]{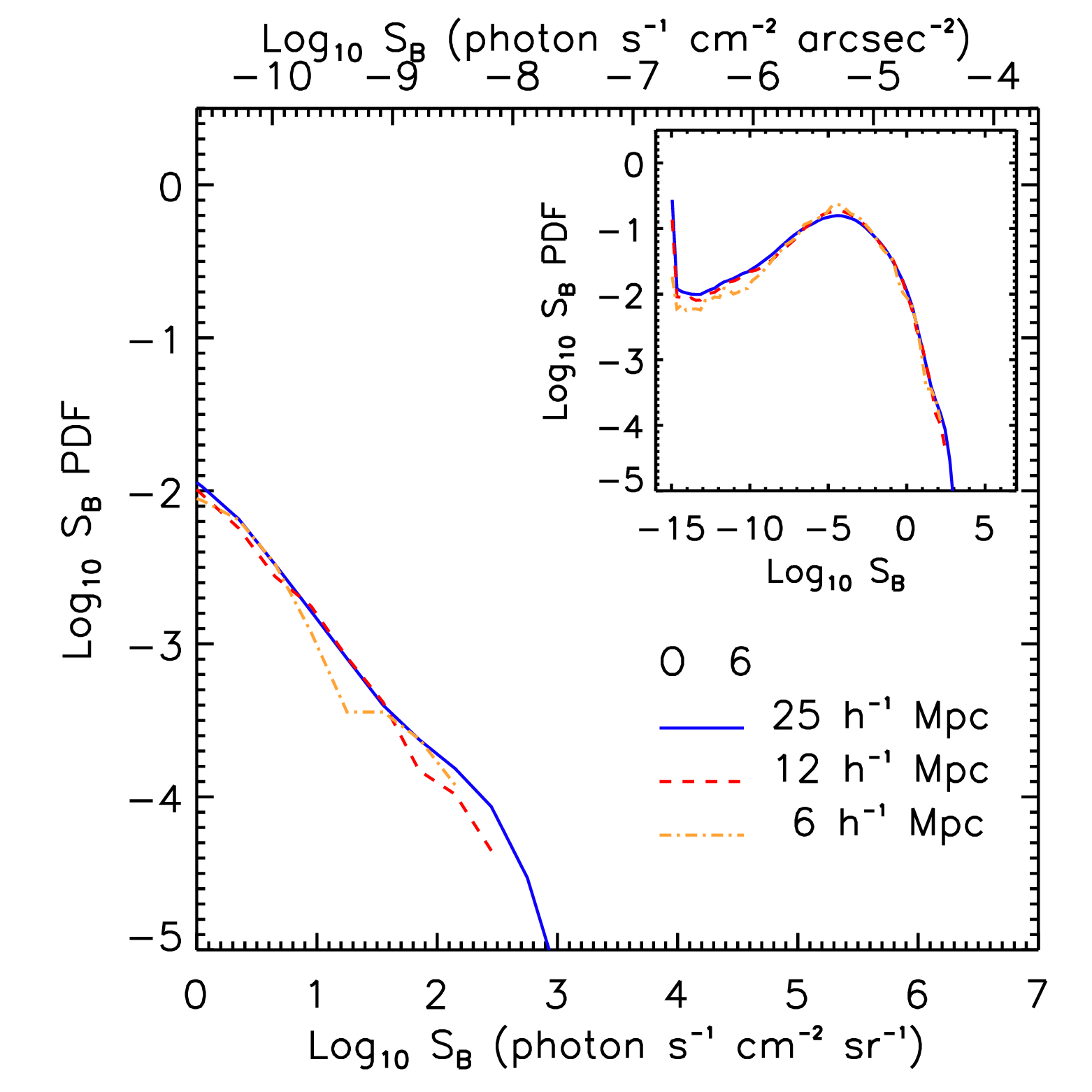}
\caption{As Fig.~\ref{lines_pdf}, but for \hil\ (left panel), \civ\ (middle panel), and \ovi\ (right panel) emission for simulations with box sizes of 25, 12.5, and 6.25 \hm\ Mpc (comoving) and a fixed numerical resolution. Different from Fig.~\ref{lines_pdf}, all maps are for 6.25 \hm\ Mpc thick slices. The results are converged with respect to the size of the simulation volume.}
\label{box_figure}
\end{figure*}

We test the effect of varying the size of the simulated box, while keeping the mass resolution constant. The runs necessarily assume different initial conditions and thus allow us to test how the large-scale structure affects the flux statistics.

In Fig.~\ref{box_figure} we compare results for \hil, \civ, and \ovi\ at $z=2$ using simulations with dark matter particle mass $m_{\rm DM} = 6.3\times 10^6$ \hm\  \msun\ and box sizes of 25, 12.5, and 6.25 \hm\ Mpc (comoving), respectively. The maps used to calculate the PDF have 2" angular resolution and 6.25 \hm\ Mpc thickness, consistent with the box size of the smallest simulation.
The results are converged with respect to the size of the simulation volume, although larger boxes sample the flux PDF to higher values.

\subsection{Slice thickness}
\label{thick}

\begin{figure*}
\centering
\includegraphics[width=0.33\textwidth]{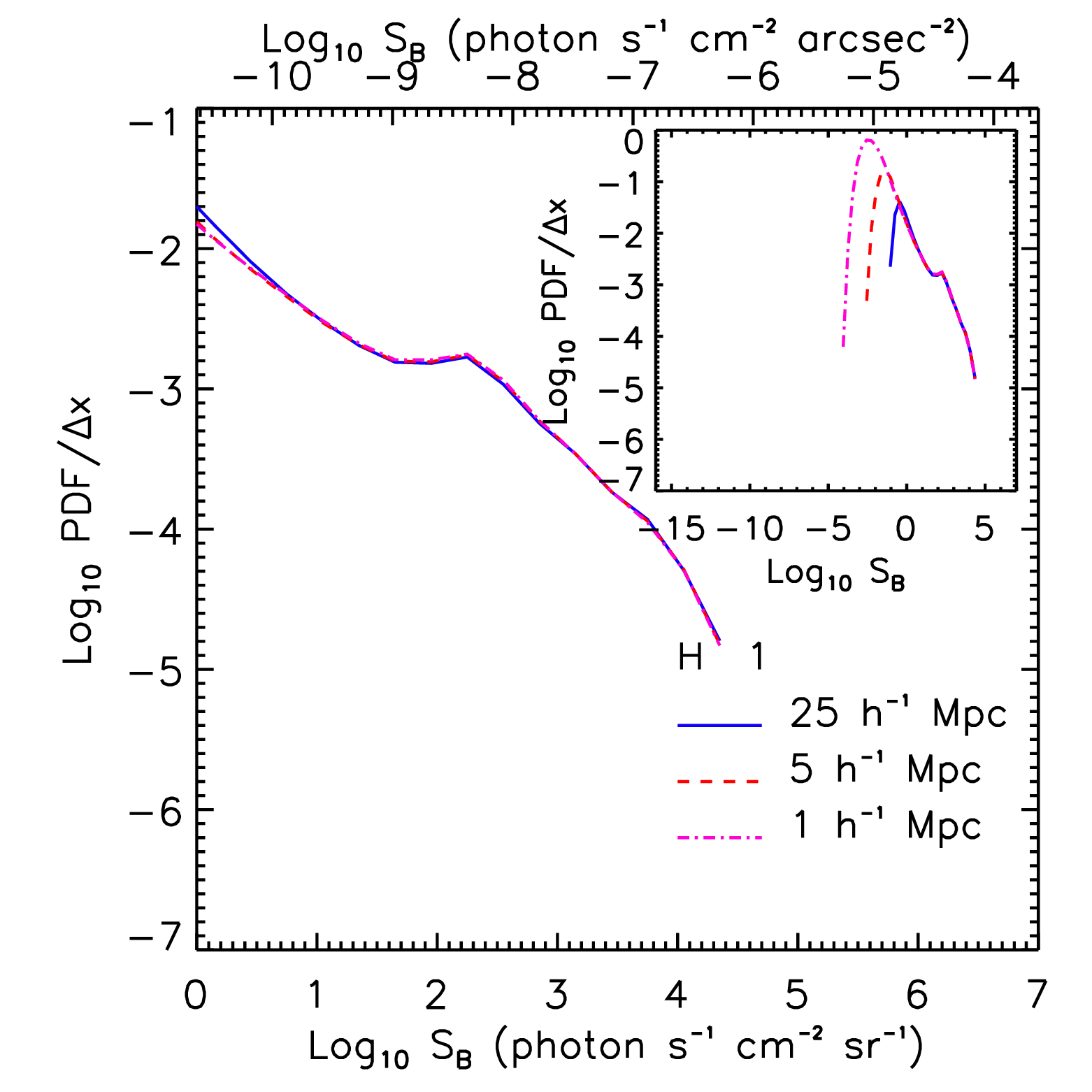}
\includegraphics[width=0.33\textwidth]{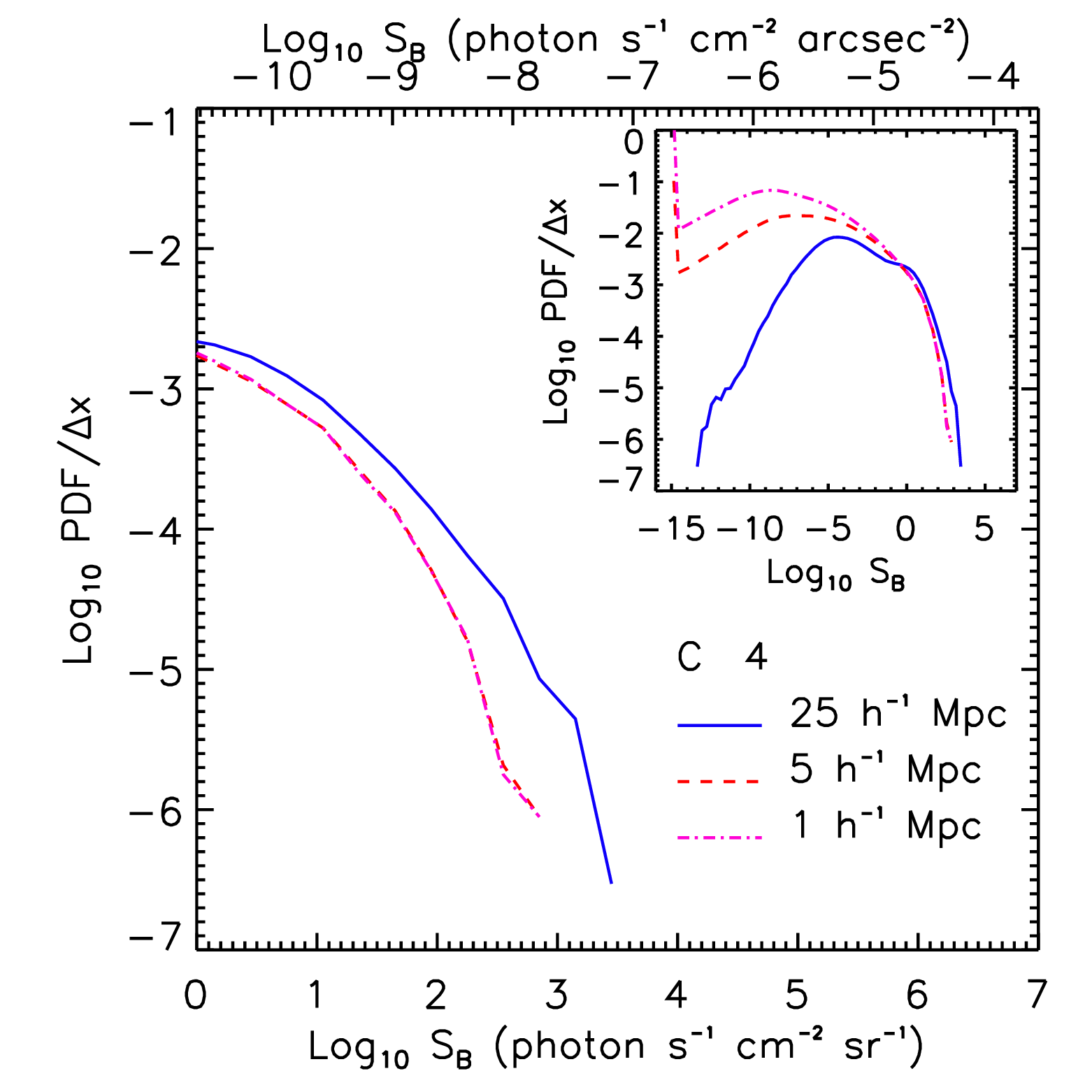}
\includegraphics[width=0.33\textwidth]{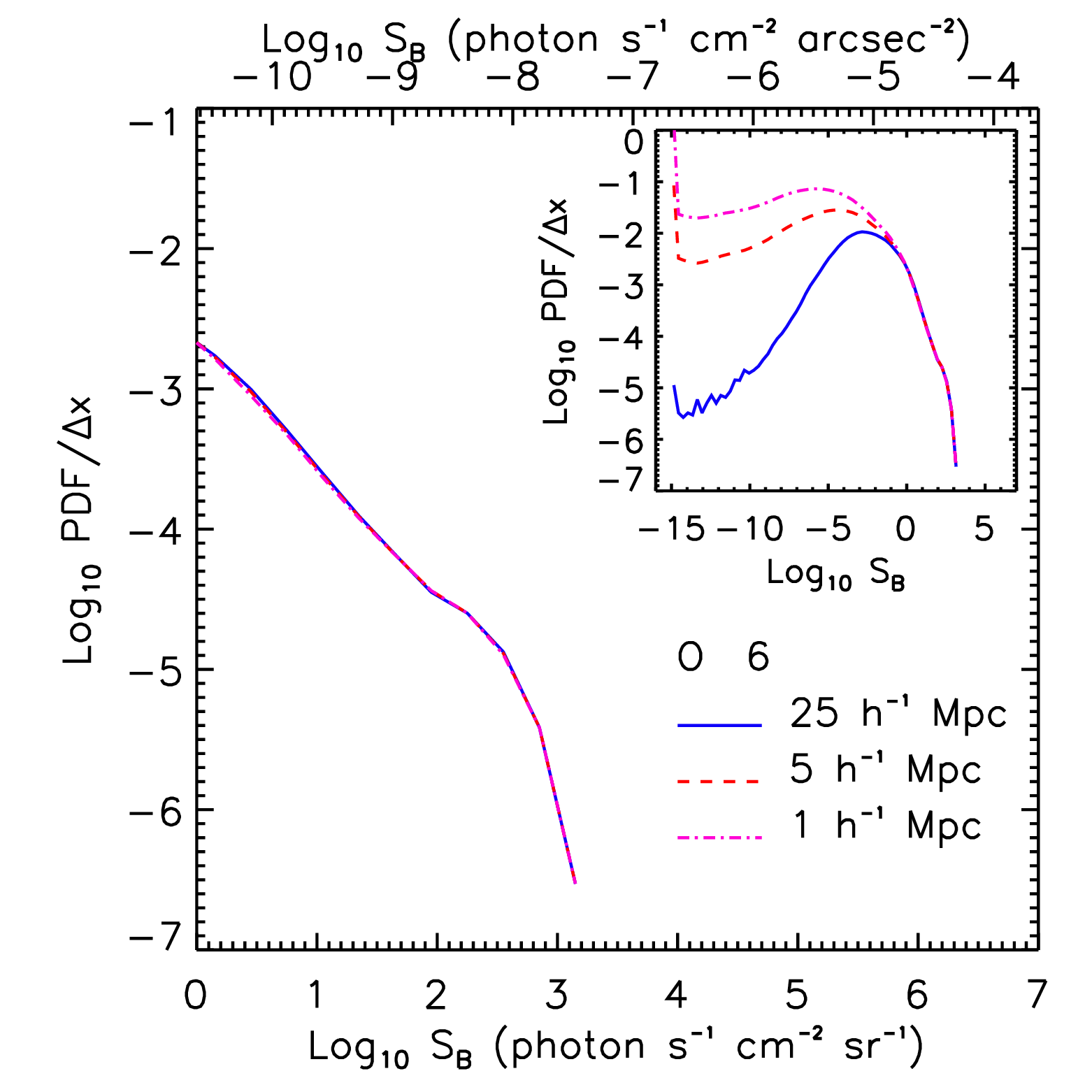}
\caption{The surface brightness PDFs for \hil\ (left panel), \civ\ (middle panel), and \ovi\ (right panel), normalised by the slice thickness $\Delta x$, for maps with varying slice thickness: 25, 5 and 1 \hm\ Mpc. The number of slices used to calculate each PDF is 1, 5 and 25 respectively. The test is for the \default\ model with box size 25 \hm\ Mpc at $z=2$. All maps assume the same angular resolution of 2". For \hil\ and \ovi\ the PDF is proportional to the slice thickness in the high flux regime. This indicates that the objects that produce the highest fluxes in the distribution are fully contained within a slice. For \civ\ there is a small difference between 5 and 25 \hm\ Mpc, but not between 1 and 5 \hm\ Mpc, indicating that projection effects slightly enhance the flux for 25 \hm\ Mpc slices.}
\label{thick_pdf}
\end{figure*}

In this Section we describe how varying the thickness of the slices we cut through the simulation box affects the shape of the flux PDF.
In Fig.~\ref{thick_pdf} we compare the $z=2$ surface brightness PDFs of, from left to right, \hil, \civ, and \ovi\ for three different, evenly-spaced slice thicknesses: 25, 5 and 1 \hm\ Mpc. To facilitate the comparison, we have normalised the surface brightness PDF by the slice thickness $\Delta x$.
Throughout this work, we have assumed a slice thickness of 25 \hm\ Mpc.

Except for \civ, the results appear to be independent of the slice thickness (within limits and in the detectable regime) when the PDF is normalised by the slice thickness. In other words, the normalisation of the PDF is proportional to the slice thickness.
This happens when the slice is thick enough to fully contain collapsed objects, but sufficiently small to avoid the superposition of multiple structures, which would strongly increase the flux in those pixels where more than one structure could be found along the line of sight. For \civ\ there is a small difference between 5 and 25 \hm\ Mpc, but not between 1 and 5 \hm\ Mpc, indicating that projection effects slightly enhance the flux for 25 \hm\ Mpc slices.

The shape of the PDF at the faint end, on the other hand, depends on the level of the flux in low-density regions and on the number of low-density structures in each slice, which is dependent on the slice thickness. As a consequence, the fraction of pixels at the lowest fluxes increases for decreasing slice thickness.

\bsp
\label{lastpage}

\end{document}